\documentclass[camera,letterpaper,nomarginnotes,nonarrowgutter]{jpaper}

\usepackage{cite}
\usepackage{amsmath,amssymb,amsfonts}
\usepackage{algorithmic}
\usepackage{graphicx}
\usepackage{textcomp}
\usepackage{xcolor}
\usepackage{fancyhdr}
\usepackage[bookmarks=true,breaklinks=true,letterpaper=true,colorlinks,citecolor=blue,linkcolor=blue,urlcolor=blue]{hyperref}

\usepackage{tabularx}
\usepackage{multirow}
\usepackage{listings}
\usepackage{booktabs}
\usepackage{makecell}
\usepackage{balance}
\usepackage{datetime}
\usepackage{datetime2}
\usepackage{setspace}

\usepackage{adjustbox}
\usepackage{multirow}
\usepackage{siunitx}
\usepackage{glossaries}
\usepackage[linesnumbered]{algorithm2e}
\setkeys{glslink}{hyper=false}

\newcommand{\affilFirst}{$^{\ast}$}

\newcommand{\githubLink}[0]{\url{https://github.com/CMU-SAFARI/Cleaning-up-the-Mess}}

\newcommand{\versionnum}[0]{1.1}

\newcommand{\secref}[1]{\S\ref{#1}}
\newcommand{\figref}[1]{Fig.~\ref{#1}}

\newcommand{\param}[1]{\textcolor{red}{#1}} % to highlight hardcoded numbers
\newcommand{\xxx}[1]{\param{XXX}} % to highlight hardcoded numbers

\newcommand{\ignore}[1]{}

\newcounter{obs}
\newcommand{\textttbreak}[1]{\lstinline[basicstyle=\ttfamily,breaklines=true]{#1}}

\definecolor{nbs}{rgb}{0.88, 0.07, 0.37}
\definecolor{agyc}{rgb}{0.37, 0.88, 0.07}
\definecolor{moegi}{rgb}{0.357, 0.537, 0.188}
\definecolor{burntorange}{rgb}{0.8, 0.33, 0.0}
\definecolor{carmine}{rgb}{0.59, 0.0, 0.09}
\definecolor{ceruleanblue}{rgb}{0.16, 0.32, 0.75}

\newif\ifcamerareadyiterations
\camerareadyiterationsfalse

\newif\ifcameraready
\camerareadytrue

\newif\ifdraft
\draftfalse

\newif\ifblind
\blindtrue

\newif\ifispassdraft
\ispassdraftfalse

\newif\ifshepherd
\shepherdfalse

\newcommand{\omthree}[1]{{#1}}

\newcommand{\omtwo}[1]{#1}

\ifcameraready
  \newcommand\nbcr[2]{#2}
  \newcommand{\atbcr}[2]{#2}
  \newcommand{\omcomment}[1]{}
  \newcommand{\nbcrcomment}[1]{}
\else 
    \newcommand\nbcr[2]{\ifnum#1=\value{version}\textcolor{red}{#2}\else{\textcolor{black}{#2}}\fi}
    \newcommand\atbcr[2]{\ifnum#1=\value{version}\textcolor{red}{#2}\else{\textcolor{black}{#2}}\fi}
    \usepackage[colorinlistoftodos,prependcaption,textsize=small]{todonotes}

    \newcommand{\omcomment}[1]{\todo[size=\scriptsize, linecolor=ceruleanblue, bordercolor=ceruleanblue, backgroundcolor=white!90!ceruleanblue]{\textcolor{ceruleanblue}{#1}}}
    \newcommand{\nbcrcomment}[1]{\todo[size=\scriptsize, linecolor=burntorange, bordercolor=burntorange, backgroundcolor=white!90!burntorange]{\textcolor{burntorange}{#1}}}
\fi

\ifshepherd
\usepackage[colorinlistoftodos,prependcaption,textsize=small]{todonotes}

\newcommand{\shepherd}[1]{\textcolor{ceruleanblue}{#1}}
\newcommand{\revision}[1]{\textcolor{burntorange}{#1}}
\newcommand{\notchanged}[1]{\textcolor{gray}{#1}}
\newcommand{\scomment}[2]{\todo[size=\scriptsize, linecolor=ceruleanblue, bordercolor=ceruleanblue, backgroundcolor=white!90!ceruleanblue]{\textcolor{ceruleanblue}{
\textbf{S#1} #2}}}
\newcommand{\rcomment}[1]{\todo[size=\scriptsize, linecolor=burntorange, bordercolor=burntorange, backgroundcolor=white!90!burntorange]{\textcolor{burntorange}{
\textbf{#1}}}}

\else

\newcommand{\shepherd}[1]{{#1}}
\newcommand{\revision}[1]{{#1}}
\newcommand{\notchanged}[1]{{#1}}
\newcommand{\scomment}[2]{}
\newcommand{\rcomment}[1]{}

\fi

\ifdraft
    \usepackage[colorinlistoftodos,prependcaption,textsize=small]{todonotes}
    \newcommand{\nbcomment}[1]{\todo[size=\scriptsize, linecolor=orange, bordercolor=orange, backgroundcolor=white]{\textcolor{nbs}{\textbf{@nb:} #1}}}
    \newcommand\nb[1]{{\color{nbs}{#1}}}
    \newcommand\agy[1]{{\color{agyc}{#1}}}
    \newcommand\atb[1]{{\textcolor{blue}{#1}}}
    \newcommand\atbnew[1]{{\textcolor{cyan}{#1}}}
    \newcommand{\mma}[1]{\textcolor{orange}{\textbf{[@mma: #1]}}}
    \newcommand\mmatodo[1]{{\textcolor{orange}{#1}}}
    \newcommand{\atbcomment}[1]{\todo[size=\scriptsize, linecolor=orange, bordercolor=orange, backgroundcolor=white]{\textcolor{blue}{\textbf{@atb:} #1}}}
    \newcommand{\hluo}[1]{{\textcolor{moegi}{#1}}}
    \newcommand{\om}[1]{{\textcolor{red}{#1}}}
    
    \newcommand{\omcr}[1]{#1}
    \newcommand{\omcrcomment}[1]{\todo[size=\scriptsize, linecolor=orange, bordercolor=orange, backgroundcolor=white]{\textcolor{blue}{\textbf{@om:} #1}}}

\else 
    \newcommand{\nbcomment}[1]{}
    \newcommand\nb[1]{#1}
    \newcommand\agy[1]{#1}
    \newcommand\atb[1]{{#1}}
    \newcommand\atbnew[1]{{#1}}
    \newcommand{\mma}[1]{}
    \newcommand\mmatodo[1]{}
    \newcommand\atbcomment[1]{}
    \newcommand\mmacomment[1]{}
    \newcommand{\hluo}[1]{{#1}}
    \newcommand{\om}[1]{{#1}}

    \newcommand{\omcr}[1]{#1}
    \newcommand{\omcrcomment}[1]{}
\fi

\ifispassdraft
    \usepackage[colorinlistoftodos,prependcaption,textsize=small]{todonotes}
    \renewcommand{\nbcomment}[1]{\todo[size=\scriptsize, linecolor=orange, bordercolor=orange, backgroundcolor=white]{\textcolor{nbs}{\textbf{@nb:} #1}}}
    \renewcommand\nb[1]{{\color{nbs}{#1}}}
\fi

\definecolor{aliceblue}{rgb}{0.94, 0.97, 1.0}

\usepackage[framemethod=tikz]{mdframed}
\newcommand{\fancycommand}[1]{%
\begin{mdframed}[backgroundcolor=aliceblue, linecolor=black, linewidth=0.8pt,skipabove=5pt]
    {#1}%
\end{mdframed}%
}

\newcommand{\atbt}[1]{\textcolor{black}{#1}}

\usepackage[most]{tcolorbox} 
\newtcolorbox[auto counter]{tkx}[2][]{%
    enhanced, breakable, center title,
    colframe = #2!45,
    colback  = #2!10,
    colbacktitle=#2!20,
    left=-0.5pt,
    right=-0.5pt,
    bottom=-2pt,
    top=-0.25pt,
    #1% 
}

\newcounter{tkw}
\newcommand\takeaway[1]{
\stepcounter{tkw}
\begin{tkx}{CornflowerBlue}
\noindent\textbf{Best Practice~\thetkw.} #1
\end{tkx}
}

\def\BibTeX{{\rm B\kern-.05em{\sc i\kern-.025em b}\kern-.08em
    T\kern-.1667em\lower.7ex\hbox{E}\kern-.125emX}}
\begin{document}

% \pagenumbering{arabic}
\title{{Cleaning up the Mess: Re-Evaluating\\the Real-System Modeling Accuracy of Ramulator~2.0}}
% \author{\normalsize{ISPASS 2026 Submission
%       \textbf{\#\submissionnumber{}}} 
%         Confidential Draft 
%         Do NOT Distribute!!
%     }
%\author{}
\author{F. Nisa Bostanc{\i}\affilFirst \quad Haocong Luo\affilFirst \quad Ataberk Olgun\affilFirst \quad Maria Makeenkova \\\omtwo{Geraldo F. de Oliveira} \quad A. Giray Ya{\u{g}}l{\i}k{\c{c}}{\i} \quad Onur Mutlu
\vspace{-4mm}
\\\\
\emph{SAFARI Research Group}
\\
\emph{ETH Z\"urich}}

% \setstretch{1.1}

\fancyhead{}
\fancyhead[C]{\textcolor{red}{CONFIDENTIAL DRAFT -- DO NOT DISTRIBUTE -- TO APPEAR IN ISPASS'26} \\ 
    \emph{\textcolor{blue}{Version \versionnum~---~\today, \ampmtime}}
}

\fancypagestyle{iterationsfirstpage}
{
    \fancyhead{}
    \fancyhead[C]{
    \textcolor{red}{CONFIDENTIAL DRAFT -- DO NOT DISTRIBUTE -- TO APPEAR IN ISPASS'26} 
    \\
    \emph{\textcolor{blue}{Version \versionnum~---~\today, \ampmtime}}
    }
}

\fancypagestyle{firstpage}
{
    \fancyhead{}
    \begin{tikzpicture}[remember picture,overlay]
    \node [xshift=150mm,yshift=-10.5mm]
    at (current page.north west) {\href{https://www.acm.org/publications/policies/artifact-review-and-badging-current}{\includegraphics[width=1.7cm]{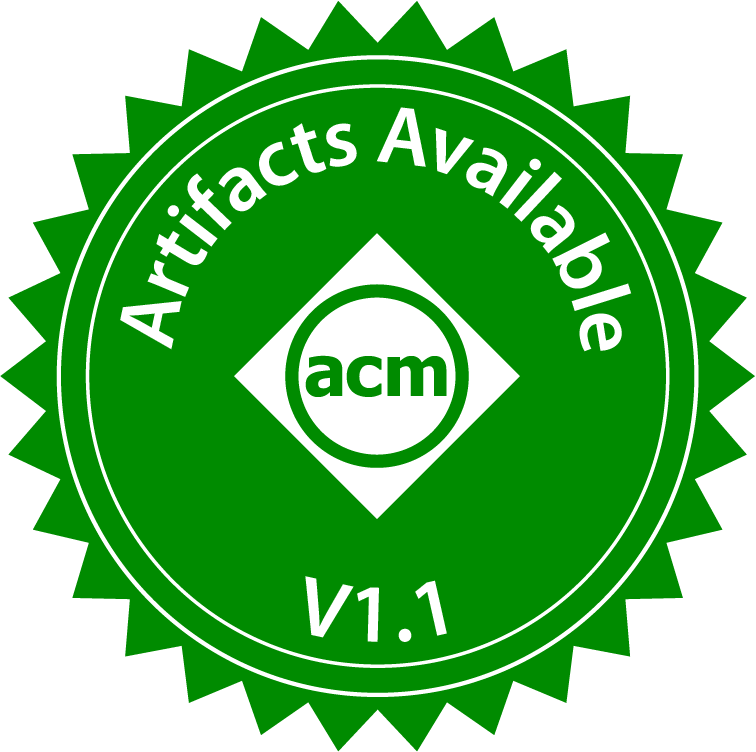}}} ;
    \node [xshift=168mm,yshift=-10.5mm]
    at (current page.north west) {\href{https://www.acm.org/publications/policies/artifact-review-and-badging-current}{\includegraphics[width=1.7cm]{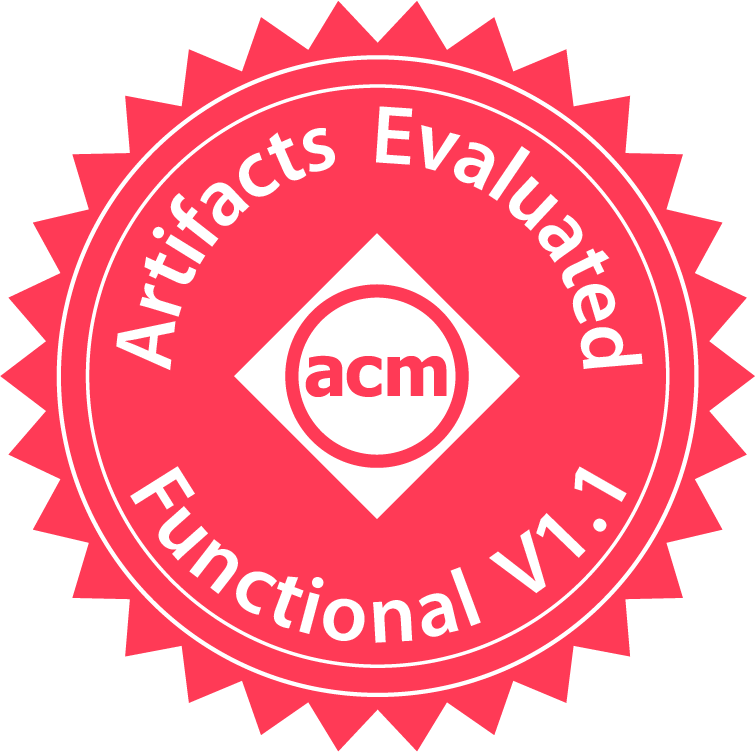}}} ;
    \node [xshift=186mm,yshift=-10.5mm]
    at (current page.north west) {\href{https://www.acm.org/publications/policies/artifact-review-and-badging-current}{\includegraphics[width=1.7cm]{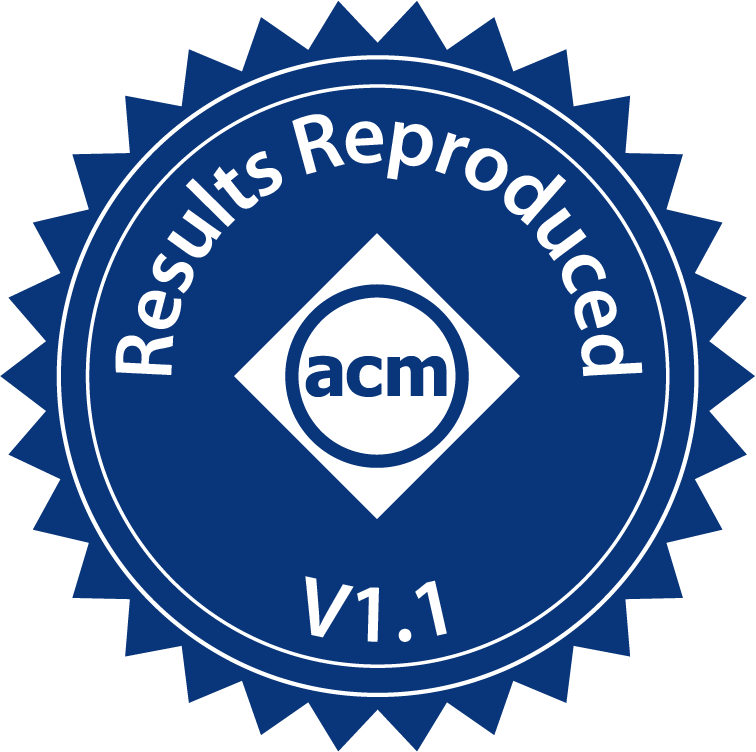}}} ;
    \end{tikzpicture}
  \renewcommand{\headrulewidth}{0pt}
  \pagenumbering{arabic}
  \fancyfoot[C]{\large\thepage}
}

\maketitle

\newcounter{version}
\setcounter{version}{6}

%Enables the camera ready header and footer
\ifcamerareadyiterations 
    \thispagestyle{iterationsfirstpage}  
    \pagestyle{plain}
    \pagenumbering{arabic}
\else
    % \renewcommand{\headrulewidth}{0pt}
    % \fancypagestyle{firstpage}{
    %     \fancyhead{} % clear all header and footer fields
    % \renewcommand{\footrulewidth}{0pt}
    % }
  \thispagestyle{firstpage}
    
\fi
% \thispagestyle{plain}
% \pagestyle{plain}

% \pagenumbering{gobble}

% \setstretch{0.99}
\renewcommand{\thefootnote}{\fnsymbol{footnote}}
\footnotetext[1]{F. Nisa Bostanc{\i}, Haocong Luo, and Ataberk Olgun are co-primary authors.}%
\renewcommand{\thefootnote}{\arabic{footnote}}
\setstretch{0.95}
\begin{abstract}
\atb{A MICRO 2024 best paper runner-up publication (the \omtwo{Mess} paper~\cite{mess}) with all
three artifact badges awarded (including ``Reproducible'') proposes a new
benchmark to evaluate real and simulated memory system performance. The
publication \omcr{contends} that Ramulator~2.0~\cite{ramulator2} and {\atb{DAMOV~\cite{damov} (ZSim+Ramulator)}} (along with other
existing memory system simulators) ``poorly resemble the actual system
performance'' \hluo{and \om{asserts that} their simulator is better}.}

\atb{In this paper, we \rcomment{D1}\revision{show that the Mess paper~\cite{mess} has 1) demonstrable technical misconfigurations, 2) methodological errors in interpreting simulation statistics, and 3) an incomplete artifact that makes {its} key results irreproducible.} \rcomment{A3}\revision{We demonstrate that the Ramulator~2.0 simulation results reported in~\cite{mess} are
{\omcr{incorrect}} due to multiple \scomment{1}{removed trivial human} configuration errors instead of inherent {simulation} inaccuracy claimed by the Mess paper.}} \hluo{We show that by correctly configuring \atb{Ramulator~2.0}, \omcr{Ramulator~2.0's} \atb{simulated} memory system performance actually resembles real system characteristics well, \atbnew{\omcr{and thus} a
key \omcr{claimed} contribution of~\cite{mess} is factually incorrect.} We also identify that the DAMOV simulation results in~\cite{mess} use wrong simulation statistics
that \omcr{are} unrelated to the simulated DRAM performance. \nb{By using correct DRAM simulation statistics, we show that DAMOV's simulated DRAM latency is not constant, in contrast to~\cite{mess}'s claim.}}
{Moreover, the \omtwo{Mess} paper's artifact repository~\cite{messzenodo, messgit} 
lacks the necessary sources (simulator code, system configurations, memory traces, etc.) to fully reproduce all \omtwo{the Mess paper's} results.}
\atbnew{We find that \omtwo{the} experiment scripts in~\cite{messzenodo,messgit} use simulator executables and other resources that are neither described in the \omtwo{Mess} paper
nor found in the artifact repository~\cite{messzenodo,messgit}}.

\ifblind
\nb{Our work 
\shepherd{identifies important issues in~\cite{mess}'s memory simulator evaluation methodology regarding Ramulator~2.0 and DAMOV. 
{We present results that validate the real-system modeling accuracy of Ramulator 2.0, and describe the reasons why \cite{mess}'s results with respect to these two simulators are incorrect.}}
%correct Ramulator~2.0 and DAMOV results that resemble the real system performance, contrary to the statements in the Mess paper}.
\scomment{2}{} We emphasize the importance of \omtwo{carefully and rigorously validating simulation results} to avoid publishing factually incorrect results and contributions.}
\else
\nb{Our work corrects~\cite{mess}'s errors regarding Ramulator~2.0 and identifies important issues in~\cite{mess}'s memory simulator evaluation methodology. We emphasize the importance of \omtwo{both carefully and rigorously validating simulation results and} contacting simulator authors and developers, in true open source spirit, to ensure these simulators are used with correct configurations and as intended. Had we, \omtwo{as developers and maintainers of Ramulator~2.0,} been \omtwo{properly} contacted before the publication of~\cite{mess}, we could have \hluo{easily} identified the errors
in the \omtwo{Mess} paper and helped the authors of~\cite{mess} to avoid publishing factually incorrect results and contributions.}\nbcomment{Double blind flag}
\fi
\atb{We \shepherd{strongly}\scomment{3}{} encourage 
%the authors of the \omtwo{Mess} paper and 
the computer architecture community
to \shepherd{consider our corrections to the Ramulator~2.0 and DAMOV results of the Mess paper~\cite{mess}}
%correct~\cite{mess}'s errors 
to prevent the propagation of inaccurate and misleading
results and to maintain the reliability of the scientific record.}
\shepherd{Our investigation also aims to stimulate discussion on artifact evaluation practices and on mechanisms for correcting results and artifacts after publication.}\scomment{4, CQ1}{}
\ifblind
\nb{To aid future works and reproduction of \emph{all} our results, we \revision{open source} \textit{all} our code and scripts.}\rcomment{D3}
\else
\nb{To aid future work, our source code and scripts are openly available at \url{https://github.com/CMU-SAFARI/ramulator2/tree/mess}.}\nbcomment{Double blind flag}
\fi
\omthree{\revision{We also discuss best practices and add sanity checks to aid users and developers of simulation tools.}}\rcomment{C1, D2}

\end{abstract}
\setstretch{1}

% \begin{IEEEkeywords}
% DRAM, memory simulation, memory benchmarking 
% \end{IEEEkeywords}

\section{Introduction}

\begin{figure*}[!t]
    \centering
    \includegraphics[width=1\linewidth]{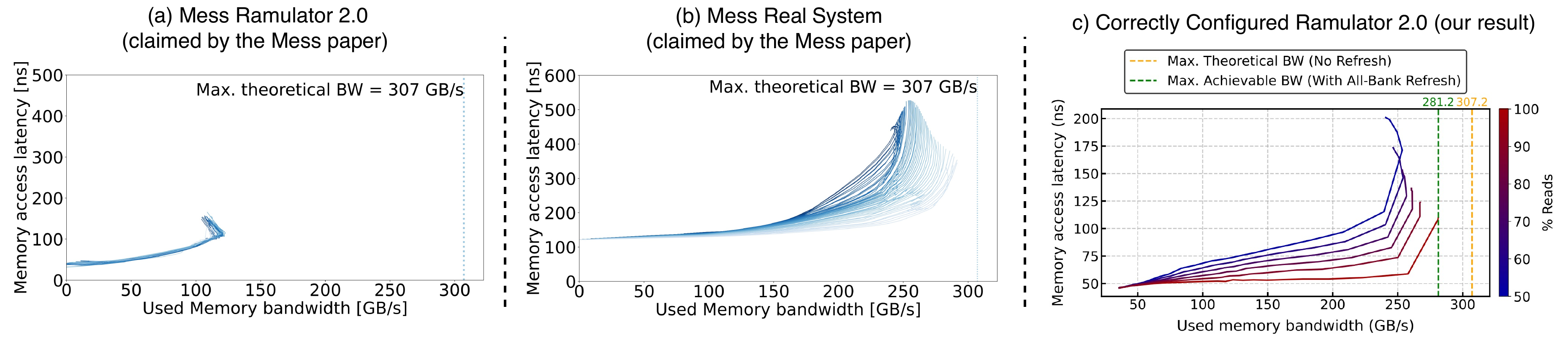}
    \caption{\omtwo{Mess} results for Ramulator~2.0 memory system simulation \omcr{from Fig. 6 in~\cite{mess}}~(a), \omtwo{Mess} results 
    for an ARM-based real system \omcr{from Fig. 4 in~\cite{mess}}~(b), our results 
    for Ramulator~2.0 memory system simulation \omcr{after correcting the errors we found in~\cite{mess}}~(c), \omtwo{as explained  in detail in Section~\ref{sec:mess_ramulator_analysis} of this paper}}
    \label{fig:comparison}
\end{figure*}

Cycle-\omcr{level} DRAM simulators enable research focused on improving the
performance, \omcr{efficiency, and robustness (including security, safety, reliability)} of DRAM-based memory systems by providing accurate \omtwo{and flexible models} for DRAM and memory controller operations. Ramulator
2.0~\cite{ramulator2,ramulator2github} \omtwo{(which builds on Ramulator~\cite{ramulator1, ramulator1github}),} is a highly modular and extensible cycle-accurate DRAM
simulator \atb{that} enable\atb{s} rapid \omcr{exploration} of \atb{new ideas in} DRAM-based memory systems. \nb{DAMOV~\cite{oliveira2021damov,damov} is a simulation framework that integrates Ramulator~\cite{ramulator1, ramulator1github} into ZSim CPU simulator~\cite{zsim} to enable fast, scalable, and cycle-level system simulation. }

A MICRO 2024 best paper runner-up publication, \emph{A \omtwo{Mess} of Memory System Benchmarking, Simulation and Application Profiling}~\cite{mess}, \omtwo{which} we refer to as ``the \omtwo{Mess}
paper'' throughout this paper, with all three artifact badges awarded (including ``Reproducible'') proposes a new benchmark to evaluate real and simulated memory system performance. \omcr{While doing so,
it makes some} strong \omtwo{negative} \omcr{claims about Ramulator~2.0 and DAMOV, which
we demonstrate to be incorrect and are due to
configuration \nb{and simulator usage} errors made in the \omtwo{Mess} paper.}

\atb{Fig.~\ref{fig:comparison}-(a) and (b) \omcr{(copied from Fig.~6 and Fig.~4 in~\cite{mess})} show average memory latency \omtwo{as a function of}
throughput (used memory bandwidth) for the \omtwo{Mess} benchmark using a Ramulator~2.0
simulation and a real system, respectively, as depicted
in~\cite{mess}.\footnote{\atb{The \omtwo{Mess} benchmark consists of two workloads
called \textit{Stream} and \textit{Pointer-Chase}. These workloads run
concurrently. The interval between subsequent memory requests in Stream is
controlled using NOPs to vary memory intensity. For example, 1K NOPs between
subsequent memory requests lead to a relatively low memory intensity and 0 NOPs
between subsequent memory requests lead to a very high memory intensity. Each
\nbcr{5}{bandwidth--latency} curve in Fig.~\ref{fig:comparison} is constructed by
connecting \omtwo{Mess} benchmark \nbcr{5}{bandwidth--latency} results in increasing memory
intensity order such that the leftmost end of the curve corresponds to the
smallest memory intensity and the other end of the curve corresponds to the
largest memory intensity \omtwo{Mess} benchmark configuration. Each curve depicts system
performance for a unique mixture of read and write memory requests in Stream
ranging from 50\% reads and 50\% writes to 100\% reads and 0\% writes.}} 
From
these two figures, the authors \omtwo{of the Mess paper} conclude that Ramulator~2.0 ``poorly resembles
the actual system performance'' \omcr{(Section I of~\cite{mess})}. They write ``The simulated memory latency is
unrealistically low and the maximum simulated memory bandwidth is only 126 GB/s
which is less than a half of the 292 GB/s measured in the actual system''
\omcr{(Section IV of~\cite{mess})}.  They
attribute the discrepancy between \omtwo{Mess} benchmark results of Ramulator~2.0
simulations and the real system to Ramulator~2.0-induced simulation errors:
``This indicates that the main source of the large simulation error is indeed
Ramulator 2'' \omcr{(Section 4 of~\cite{mess})}.}
\atbnew{The authors also show other \omtwo{
%odd and 
unexpected}\scomment{5}{removed odd} results. For example, the DRAM latency results from DAMOV (ZSim+Ramulator)~\cite{damov} \omcr{(depicted in Section 4 of~\cite{mess})} are unrealistically low 
and \omtwo{they seem to} remain constant regardless of system load.} 

\atb{In this paper, we show that the Ramulator~2.0 simulation results in the \omtwo{Mess} paper are
\emph{\omtwo{incorrect}} and, at the time of the
publication of~\cite{mess}, \emph{irreproducible}. Fig.~\ref{fig:comparison}-(c) shows
our reproduction of the memory \nbcr{5}{bandwidth--latency} curves for the \omtwo{Mess} benchmark using the}
\atb{open source version of Ramulator~2.0~\cite{ramulator2} \omtwo{after correcting the error we found in~\cite{mess}}.
\ifblind
\else
\footnote{\atb{We
freely and openly release all sources and data used in this work
at~{{\url{https://github.com/CMU-SAFARI/ramulator2/tree/mess}}}. \omcr{One can reproduce \emph{all} our graphs using the scripts we provide in that repository.}}}
\fi
These results clearly show that
Ramulator~2.0 does \emph{not} poorly resemble real system performance contrary
to what the \omtwo{Mess} paper claims.}

{We carefully investigate the \omtwo{Mess} paper's open source
artifacts~\cite{messgit,messzenodo} and \omcr{have} communicated with \omcr{the \omtwo{Mess} paper's} authors
to identify the root causes of the errors in~\cite{mess}.\nbcomment{Communication mentioned} We identify
{two} major Ramulator~2.0 configuration errors and {\omcr{two}} other scholarship issues regarding Ramulator~2.0 in the \omtwo{Mess} paper. 
\atb{We describe these issues in detail
in Sections~\ref{sec:mess-paper-errors} and~\ref{sec:mess-scholarship-issues}.}}
\atbnew{In short, the Ramulator~2.0 results in the \omtwo{Mess} paper are subject to
%gross 
\scomment{6}{removed gross} configuration errors. Therefore, these results
do \emph{not} support the 
% strong
\omtwo{negative}
claims (asserted in the \omtwo{Mess} paper) about Ramulator~2.0's 
modeling \scomment{7}{removed redundancy}accuracy.\footnote{\omtwo{Strong negative} \atbnew{assertions made in the \omtwo{Mess} paper such as 
\shepherd{``... the most complex and
trusted memory model shows the highest simulation error...'' \omtwo{(Section 4 of~\cite{mess})}} are incorrect
because the source of the error is the misconfiguration and misuse of Ramulator~2.0 in the \omtwo{Mess} paper.}} 
\shepherd{{Ramulator 2.0 can be correctly configured}
%The correct configuration can be determined 
by careful and rigorous validation and/or by contacting Ramulator~2.0 authors and developers after observing unexpected results, to ensure that the simulator is used correctly and as intended.\footnote{\notchanged{The Mess paper's authors verified via email exchanges that they intentionally did \emph{not} contact the developers of the memory system simulators evaluated in the Mess paper before publication of the paper.}}}}\scomment{8,9}{}

We also identify that the reason for unrealistically low
and \omtwo{seemingly} constant DRAM latency results for DAMOV from the \omtwo{Mess}
paper
is that the \omtwo{Mess} paper's authors \omcr{report} simulator statistics that are
\emph{not} updated during the DRAM simulation. 
{Moreover, the \omtwo{Mess} paper's artifact 
repository~\cite{messzenodo, messgit} does \emph{not}
contain the relevant DRAM simulation statistics \omcr{from Ramulator} for
DAMOV.} \atb{We find that \omtwo{the experiment scripts of the Mess paper} use simulator executables and other resources that are neither described in the \omtwo{Mess} paper
nor found in the \omtwo{Mess} paper's artifact repository~\cite{messgit,messzenodo}.}
\nb{We independently evaluate DAMOV's simulated memory system performance using the open source version of DAMOV~\cite{damov}. We show that, when using the correct DRAM simulation statistics from Ramulator, 1)~the simulated DRAM latency does \textit{not} remain constant in contrast to the results claimed by the Mess paper~\cite{mess}, and 2)~DAMOV's \nbcr{5}{bandwidth--latency} curves resemble real system characteristics.}

{Our work \shepherd{provides corrections to the Ramulator~2.0 and DAMOV results of the Mess paper~\cite{mess}}\scomment{10}{}
and identifies important scholarship issues in the memory simulator evaluation methodology \nb{of the Mess paper~\cite{mess}}. \notchanged{We emphasize the importance of contacting simulator authors and developers\nbcomment{contacting developers mentioned} \omcr{(especially before publishing)}, in true open source spirit, to ensure these simulators are used with correct configurations and as intended.}}
\atb{We \shepherd{strongly}\scomment{11}{} encourage 
the computer architecture community
to \shepherd{consider our corrections to the Ramulator~2.0 and DAMOV results of the Mess paper~\cite{mess}}
to prevent the propagation of inaccurate and misleading
results and to maintain the reliability of the scientific record.}
\omthree{\shepherd{Our investigation also aims to stimulate discussion on artifact evaluation practices and on mechanisms for correcting results and artifacts after publication.}}\rcomment{CQ2}
\revision{We discuss four best practices and {add} sanity checks 
to aid {users and developers} \omthree{of simulation tools} (\S\ref{sec:discussion}).}\rcomment{C1, D2}
\nbcr{2}{To aid future works and reproduction of all our results, we open source all our code and scripts at \githubLink{}.}

\section{\omcr{Analysis of the} Evaluation of Ramulator~2.0\\in the \omtwo{Mess} Paper}
\label{sec:mess_ramulator_analysis}
\subsection{The \omtwo{Mess} Benchmark}
The \omtwo{Mess} benchmark aims to characterize a memory system based on a set of
\nbcr{5}{bandwidth--latency} curves.
To this end, \atbnew{the \omtwo{Mess} paper} introduces
the \omtwo{Mess} benchmark. The \omtwo{Mess} benchmark is made up of two concurrently running
workloads, \textit{Stream} and \textit{Pointer
Chase}~\cite{verdejo2017microbenchmarks, verdejo2018main}. Each curve is given
by running these workloads with a pre-configured \omcr{fraction} of read and write
requests for the Stream, ranging from 50\% reads and 50\% writes to 100\% reads and 0\% writes.
Each point on a curve is given by a run of the benchmark with varying memory
intensity. This is achieved by varying the interval between subsequent Stream
requests by inserting NOPs. For example, 1K NOPs between subsequent memory
requests lead to a relatively low memory intensity, and 0 NOPs between
subsequent memory requests lead to a very high memory intensity.

\subsection{\omcr{\omtwo{Mess} Paper's~\cite{mess} Ramulator~2.0 Results}}
\begin{figure}[!h]
    \centering
    \includegraphics[width=1\linewidth]{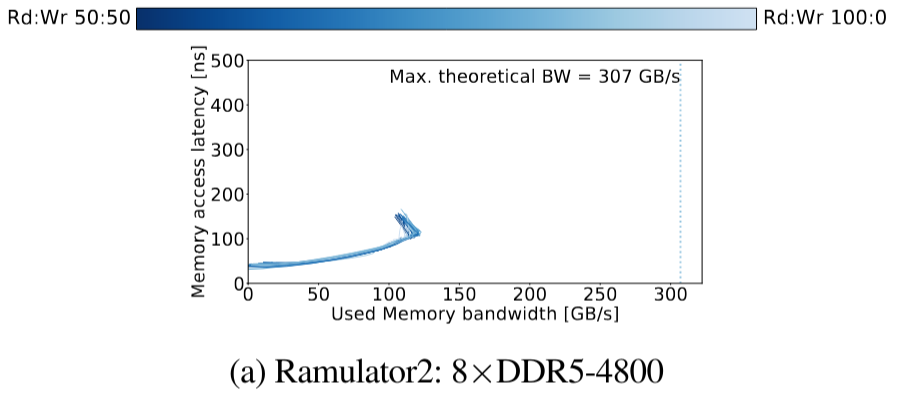}
    \caption{\atb{\nbcr{5}{Bandwidth--latency}} curves for Ramulator~2.0 \agy{(copied from Fig. 6 in~\cite{mess})}}
    %\atb{as depicted in \omcr{Fig. 6 in~\cite{mess}}}
    \label{fig:mess_ram2}
\end{figure}

Fig.~\ref{fig:mess_ram2} \atb{(Fig.~\ref{fig:comparison}-a plotted again 
for the reader's convenience)} 
shows the \nb{\nbcr{5}{bandwidth--latency} curves presented
in~\cite{mess} \omcr{(Fig. 6 in~\cite{mess})} for} Ramulator~2.0 \nb{obtained with} the \omtwo{Mess} benchmark. This
result \atb{unexpectedly} \omtwo{and oddly} shows that the simulated memory system 
\atb{1)~uses significantly smaller bandwidth
than what is available (depicted as the ``Max. theoretical BW'' in the figure)} 
and 2)~yields a relatively small memory access latency. 

\revision{\subsection{Reproducibility Issues in \cite{mess}}}\rcomment{D1}

\atb{We note that it \omtwo{was} 
%\emph{impossible} 
\shepherd{\textit{not} possible}\scomment{12}{removed impossible}
to reproduce \agy{\emph{any}} Ramulator~2.0 result
from the \omtwo{Mess} paper artifacts available at the time of \atbnew{the} publication and artifact
evaluation \atbnew{of the \omtwo{Mess} paper}~\cite{messgit,messzenodo}. Thus, we contacted the 
authors \omcr{of~\cite{mess} on 16 October 2024} to obtain
1)~the Ramulator~2.0 source code and 2)~the \omtwo{Mess} benchmark memory traces used in
Ramulator~2.0 simulations reported in~\cite{mess}. The missing sources needed to replicate the \omtwo{Mess}
paper's Ramulator~2.0 results \omcr{were} \omtwo{eventually} uploaded to \nb{the} \omtwo{Mess} GitHub
repository~\cite{messgit} following our contact \omcr{on 18 July 2025},
\emph{10 months after} the \omtwo{Mess}
paper \omcr{was} awarded \omcr{(on 8 September 2024)} all three artifact evaluation badges.}

\subsection{Fundamental Configuration Errors in~\cite{mess}}
\label{sec:mess-paper-errors}

\atb{We identify two fundamental configuration errors in the \omtwo{Mess} paper's Ramulator~2.0 simulations. We attribute the discrepancy in bandwidth and memory access latency between
the \omtwo{Mess} paper's Ramulator~2.0 simulations and real system evaluation to these fundamental errors.}
\atbnew{We also identify \shepherd{two}\scomment{14}{} other scholarship issues.}

\atbnew{First, the \omcr{\omtwo{Mess} paper's} \omtwo{scripts} \omcr{(in the newly-uploaded source code \omtwo{on 18 July 2025})} instantiate and use a single-channel DDR5 Ramulator~2.0 memory system. 
The authors extrapolate single-channel DDR5 \revision{bandwidth measurements}\rcomment{C2}
%results 
to 16-channel DDR5 
\revision{bandwidth measurements} 
%results 
by multiplying the observed bandwidth by 8 instead of 16.
This is because they \om{\omcr{incorrectly}} assume that a DDR5 
channel in Ramulator~2.0
is \omcr{64 bits wide}, consisting of two independent 32-bit channels. However, the source code of Ramulator~2.0 clearly shows that a DDR5 channel is \omcr{32 bits wide}\footnote{The \texttt{m\_channel\_width} parameter in DDR5.cpp in the Ramulator~2.0 repository~\cite{ramulator2github}.} in Ramulator~2.0\revision{, which\rcomment{C2} is consistent with JEDEC DDR5 standards~\cite{jedecddr5c, jedecddr5udimm, jedecddr5rdimm}}.\footnote{\revision{The JEDEC DDR5 SDRAM component~\cite{jedecddr5c} and UDIMM module~\cite{jedecddr5udimm} standards consistently \omthree{use} the term ``channel''. The JEDEC DDR5 RDIMM standard~\cite{jedecddr5rdimm} uses the terms ``channel'' and ``sub-channel'' interchangeably. However, both ``channel'' and ``sub-channel'' in these JEDEC \omthree{DDR5} documents~\cite{jedecddr5c, jedecddr5udimm, jedecddr5rdimm} are 32-bit wide.}}
The \omtwo{Mess} paper
misleadingly presents (extrapolated) 8-channel (32-bit each) DDR5 results as 16-channel (32-bit each) DDR5 results, 
leading to \omcr{incorrect} claims like ``...the maximum simulated memory bandwidth is only
126 GB/s which is less than a half of the 292 GB/s measured
in the actual system'' \omtwo{in Section 4 of~\cite{mess}.} The maximum bandwidth
of \omcr{an} 8-channel DDR5-4800~\cite{jedecddr5c} is 153.6 GB/s~\cite{jedecddr5c}, \emph{not} 307 GB/s. Thus, a bandwidth of 
more than 153.6 GB/s should \emph{not} be expected
of an 8-channel DDR5-4800 configuration. We show that a correctly configured
16-channel DDR5-4800 Ramulator~2.0 simulation achieves 281.1 GB/s bandwidth
(\omcr{as depicted in Fig.~\ref{fig:comparison}-c;} see Section~\ref{sec:results} \omcr{for a detailed description}).}

{Second, the \omtwo{Mess} paper's \omtwo{scripts} use the SimpleO3 frontend of Ramulator~2.0~\cite{ramulator2github} 
with unrealistically low latency configurations compared to a real system.} 
For example, they configure 1) the cache latency to be
\emph{0} CPU clock cycles and 2) each CPU core to have
1024 miss-status holding registers (MSHRs). Compared
to a real CPU, such a configuration has 1) significantly smaller round-trip latency between a core and DRAM, \nbcr{4}{and} 2)
significantly lower queuing delay for memory requests when memory intensity is high.
\rcomment{C4}
\omthree{\revision{About} Ramulator~2.0's simulation accuracy,} \revision{the Mess paper} claims that ``...the simulated
memory latency is unrealistically low...'' \omtwo{(Section 4 of~\cite{mess})}.
\revision{Because the unrealistic cache setup is in part responsible for the low 
latency observed in~\cite{mess}, 
the cache setup in~\cite{mess} constitutes a fundamental configuration error. Thus, the 
Ramulator 2.0 simulation accuracy claim made in the Mess paper~\cite{mess} is {incorrect}.}

\subsection{Scholarship Issues in~\cite{mess}}
\label{sec:mess-scholarship-issues}

\nb{First, }
\omcr{the \omtwo{Mess paper} do\omtwo{es} \emph{not} disclose the unrealistically low latency configurations of the SimpleO3 frontend of Ramulator~2.0 in the paper. The Ramulator~2.0 configuration file available in the artifact repository also does \emph{not} include any of these configurations. After the \omtwo{Mess} paper's authors uploaded their Ramulator~2.0 source code,\scomment{13}{removed 10 months after statement}
% ,
%\shepherd{following our contact}\scomment{13}{revised}, 
we \omtwo{found (via email exchanges with the Mess paper's authors)} that they implement the unrealistically low latency configurations by \emph{directly modifying the source code of Ramulator 2.0}, without \nbcr{3}{any documented changes}.}

\nb{Second, }
\atb{the \omtwo{Mess paper} uses \emph{different} \omtwo{Mess} benchmark methodologies for real
systems and simulators \emph{without} disclosing this difference in the paper.\footnote{\omtwo{Identified via email exchanges with the Mess paper's
authors.}}
For real systems \omcr{(e.g., Fig. 5-a in~\cite{mess})}, the \omtwo{Mess} benchmark is run as described in the \omtwo{Mess} paper
\nb{(i.e., by concurrently running \textit{pointer-chasing} and \textit{Stream}
workloads and measuring the memory latency of the \textit{pointer-chasing}
workload's memory accesses)}. For simulators \omcr{(e.g., Figures 5-b, -c, -d, -e, and -f in~\cite{mess})}, the authors \emph{do not run a
pointer-chasing workload} and \omtwo{they} measure the average memory latency for \emph{\omcr{only}} the
\emph{Stream} workload. 
\omcr{Thus,} the \omtwo{Mess} paper's authors compare apples to oranges \atbnew{by running \omtwo{and measuring} different workloads in simulators and in real systems.} 
\omcr{And, this} difference
in simulation and real system evaluation methodologies is \emph{not} disclosed in the \omtwo{Mess} paper}.

\atb{Through our comprehensive investigation of~\cite{mess}'s Ramulator~2.0
artifacts, we conclude that the Ramulator~2.0 results in~\cite{mess} are subject
to \omtwo{multiple}\scomment{15}{removed "serious"}
%serious
configuration and methodological errors that could have been}
avoided by careful \omcr{and rigorous} validation.

\section{\atb{Our} Evaluation Methodology}
\label{sec:evaluation_ramulator}

\atb{To show that Ramulator~2.0, when configured \omcr{properly} and used correctly, yields
{reasonable} \omcr{and realistic} \nbcr{5}{bandwidth--latency} curves, we evaluate a modern DDR5 memory
system in \omcr{Ramulator~2.0 using the same access patterns as described in the \omtwo{Mess} benchmark.}}

\nb{We} introduce a new Ramulator~2.0 frontend module, the \textit{\omtwo{Mess} Request
Generator,} \omcr{which we open source, along with all our
code and scripts \nbcr{2}{in our repository~\cite{self.github,cutmess.github}} dedicated to this work}.
\omtwo{Our} \omtwo{Mess} Request Generator frontend is designed to emulate the \atbnew{memory access patterns of the} \omtwo{Mess} benchmark by
generating a mixture of \omtwo{random access} (\textit{random}) and \omtwo{streaming} (\textit{stream}) read \nb{requests}.
\atb{The request generator issues \textit{random}} requests sequentially \atb{(to emulate a
pointer-chasing workload)}, meaning that \atb{the request generator issues a}
new random request \emph{only} when the previous \atb{\textit{random} request returns from the memory controller}.
\atb{The request generator issues a configurable mix of} \textit{stream}
\atb{read} and \atb{\textit{stream} write} requests \atb{to the memory
controller} as often as every \atb{DRAM command interface clock} cycle
\atb{(e.g., \SI{0.416}{\nano\second} for DDR5-4800~\cite{jedecddr5c})}. \atb{We modulate the
frequency of \textit{stream} requests using NOPs \omtwo{between two consecutive requests}.} The sequence of \textit{stream}
requests are designed to take advantage of bank-group- and bank-level parallelism and
row hits \atb{to fully utilize DDR5 data transfer bandwidth}. \rcomment{C3}\revision{To do so, we use a \textit{stream} access pattern that overlaps the latency of memory requests by issuing requests that are interleaved across bank-groups and banks in each channel.} 
The request generator keeps issuing \textit{stream} requests
while waiting for a \textit{random} read request to return from the memory controller.

\noindent
\atb{\textbf{Read/Write Mix.}} The \omtwo{Mess} Request Generator frontend for Ramulator~2.0 is configurable with a read ratio between 0.5 and 1,
which is applied to the stream memory requests. A read ratio of 1 corresponds to
streaming accesses consisting of 100\% reads, while a read ratio of 0.5 yields 50\%
read and 50\% write requests for the streaming memory accesses. \atb{We use
read ratios of 0.5, 0.6, 0.7, 0.8, 0.9, and 1.0 in our evaluation.}

\noindent
\atb{\textbf{\textit{Stream} Bandwidth Use Modulation.}} To simulate various
system loads, we further extend the \omtwo{Mess} Request \nbcr{4}{G}enerator with a configurable
NOP frequency. In every Ramulator~2.0 frontend cycle, the \omtwo{Mess} Request \nbcr{4}{G}enerator can either
issue a memory request or not, depending on the value of the NOP frequency. For
example, \omtwo{with} a NOP frequency of 1, every cycle leads to a memory request being
issued, while a NOP frequency of 100 means that every 100 simulated cycles, a
memory request is issued.

We use the \omtwo{Mess} Request Generator frontend for Ramulator~2.0 to construct the \nbcr{5}{bandwidth--latency}
curves. We  
use the \atb{Ramulator~2.0 memory system} configuration in Table~\ref{tab:system_config}.
We \revision{open source}\rcomment{D3} \omtwo{our} Ramulator~2.0 configuration along with all source \omtwo{code} and scripts to reproduce our
results \nbcr{2}{in~\cite{self.github, cutmess.github}}.\footnote{\atb{Our Ramulator~2.0 \nbcr{2}{codebase~\cite{self.github, cutmess.github}} dedicated for this work introduces \emph{only} minimal
and necessary changes to the open source Ramulator~2.0 repository~\cite{ramulator2github}. First, we implement
the \omtwo{Mess} Request Generator frontend we describe. Second, we add the timing parameters for the DDR5-4800AN standard~\cite{jedecddr5c}
to remain consistent with what is evaluated in the \omtwo{Mess} paper~\cite{mess}. Third, we disable write-to-read
forwarding in the memory controller \omcr{to ensure all memory requests are served by DRAM}. Fourth, we add utility functions 
to record various statistics about memory requests and to programmatically
configure the \omtwo{Mess} Request Generator.}}

\vspace{1mm}
\begin{table}[ht]
\centering
\caption{Simulated System Configuration}
\begin{tabularx}{\linewidth}{lp{5cm}}
\toprule
\textbf{Frontend:} & \omtwo{Mess} Request Generator for Ramulator~2.0~\cite{ramulator2}\\
\midrule
\textbf{DRAM:} & DDR5-4800AN~\cite{jedecddr5c}; 16~channels \revision{(32-bit each)}, 1~rank, 8~bank~groups, 4~banks\\
\midrule
\makecell[l]{\textbf{Timing} \textbf{Parameters:}} & $t_{RCD}$: \SI{14.166}{\nano\second}, $t_{RAS}$: \SI{32.000}{\nano\second}, $t_{RP}$: \SI{14.166}{\nano\second}, $t_{RTP}$: \SI{7.5}{\nano\second},  $t_{CCD\_S}$: 8 nCK, $t_{RFC}$: \SI{295}{\nano\second}, $t_{REFI}$: \SI{3.9}{\micro\second}\\
\midrule
\makecell[l]{\textbf{Memory} \textbf{System:}} & FR-FCFS request scheduling policy~\cite{scott2000memory, zuravleff1997controller}, All-bank refresh, Read queue size:~32, Write queue size:~32 \\
\midrule
\end{tabularx}
\label{tab:system_config}
\end{table}
\rcomment{C2}

To
construct each \nbcr{5}{bandwidth--latency} curve, we run the simulation for \atb{20K}
\textit{random} read requests and measure \emph{only} the latency of the \textit{random} read
requests \omcr{(i.e., same as in the \omtwo{Mess} benchmark)}. \atb{We sweep NOP frequency from 1 to 10000 to vary system load.}

\section{\atb{Key Results}}
\label{sec:results}

\begin{figure}[!h]
    \centering
    \includegraphics[width=1\linewidth]{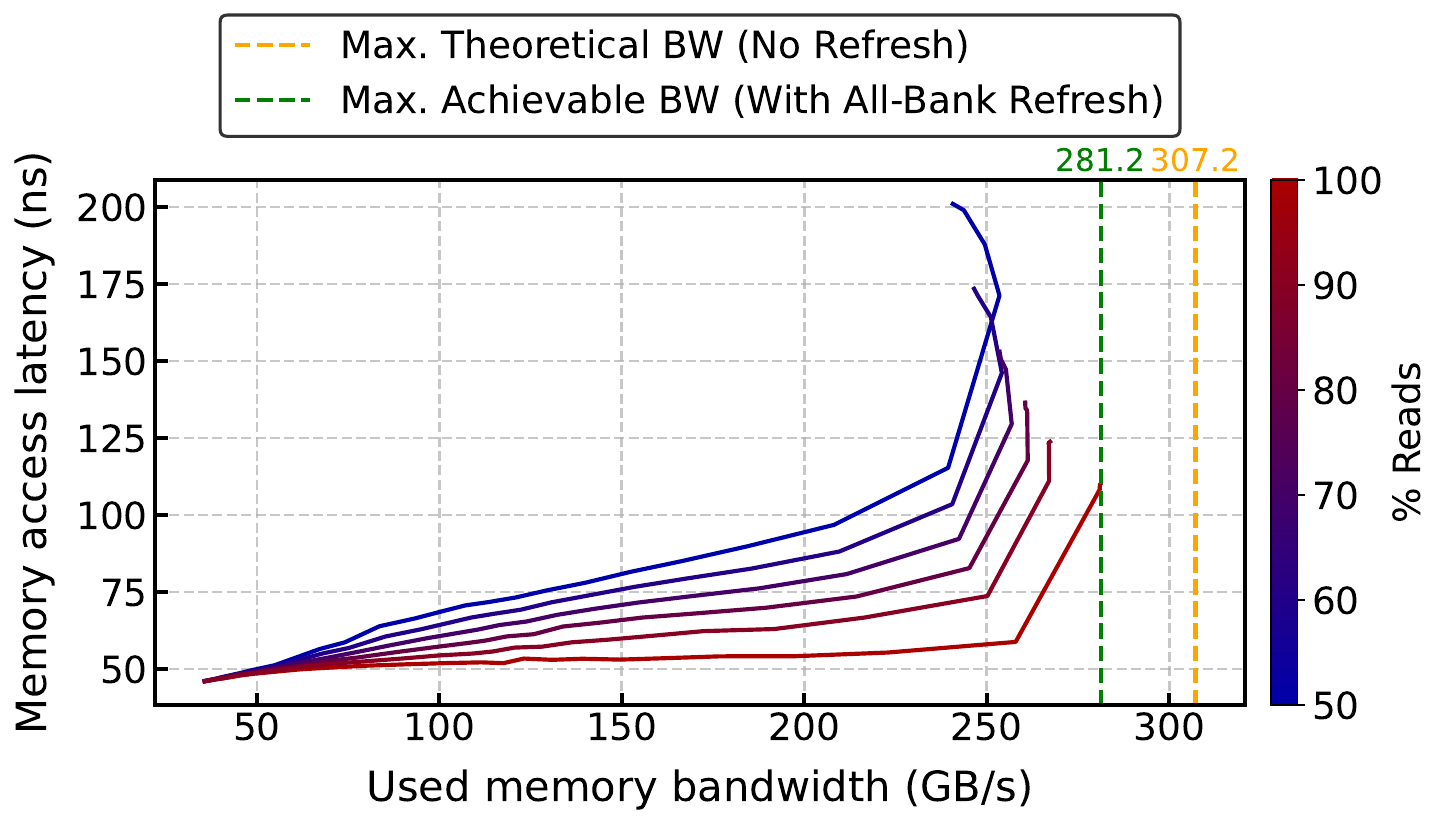}
    \caption{\nbcr{5}{Bandwidth--latency} curves for DDR5-4800AN in Ramulator~2.0 using the \omtwo{Mess} Request Generator frontend \omcr{and configuration in Table~\ref{tab:system_config}}}
    \label{fig:main_res}
\end{figure}

Fig.~\ref{fig:main_res} shows the \nbcr{5}{bandwidth--latency} curves generated 
using Ramulator~2.0 with the \omtwo{Mess} Request Generator frontend and the 
configuration given by Table~\ref{tab:system_config}.
\atb{The maximum theoretical bandwidth for 
16 channels of DDR5-4800 (indicated by the orange line in the figure) is 307.\atb{2} GB/s \omcr{(as calculated by 16$\times{}$19.2 GB/s~\cite{jedecddr5c})}.}
\atb{The maximum bandwidth \emph{cannot} be sustained during standard 
operation due to \emph{all-bank periodic refresh operations}~\cite{jedecddr5c} that \omcr{conflict with and delay} 
data transfers. The memory controller can sustain the maximum bandwidth as long as
subsequent $READ$/$WRITE$ commands are issued at the end of every short
column-to-column delay ($tCCDS$). To serve a periodic refresh operation,
the memory controller must precharge the open bank and issue a refresh command.
Doing so introduces a minimum timing penalty of the sum of read-to-precharge ($tRTP$), 
precharge latency ($tRP$), 
refresh latency ($tRFC$), and activation latency ($tRCD$) between two subsequent $READ$ commands.}

\omcr{Taking such refresh operations and their impact on data transfers into account,} we compute the \emph{maximum achievable bandwidth}
from the maximum theoretical bandwidth
as 281.2 GB/s (indicated by the green line in the figure) from the following equations: 
$$BW_{achievable} = BW_{theoretical} \cdot \left( 1 - \frac{REF_{penalty}}{tREFI}\right)$$
$$REF_{penalty} = tRTP + tRP + tRFC + tRCD$$

\atb{We make {three} major observations from Fig.~\ref{fig:main_res}. 
First, we observe that
the maximum achieved memory bandwidth (281.1 GB/s for \omcr{a read ratio of 1, i.e.,} 100\% reads, and NOP \omcr{frequency of 1, i.e., no NOP cycles}) 
\omcr{almost exactly} matches the maximum achievable memory bandwidth. We thoroughly investigate
the DRAM command sequence modeled by Ramulator~2.0 to understand the small
difference between the maximum achieved \omtwo{bandwidth} \omcr{(281.1 GB/s)} and \omtwo{the maximum} 
achievable \omtwo{bandwidth} \omcr{(281.2 GB/s)} values. We observe that
the $REF_{penalty}$ slightly varies for each periodic refresh operation. \omcr{For example,} if the memory
controller opens a DRAM row to serve a random access read request and schedules a periodic
refresh operation immediately after this read request, the delay between
two $READ$ commands interrupted by a refresh operation is 
slightly higher \omcr{than $tRTP + tRP + tRC + tRCD$}: $(tRAS - tRCD) + tRP + tRC + tRCD$.}

\atb{Second, as the fraction of write memory requests increases, the \nbcr{5}{bandwidth--latency} 
curves are curved more aggressively
towards the top left corner of the figure because a larger fraction of
write memory requests induces a higher latency and bandwidth penalty
due to more frequently incurred DRAM bus turnaround 
latency~\cite{lee2010dram}.}

\atb{Third, we observe that the latency of the random access memory requests
consistently increases with used memory bandwidth because interference 
from \textit{stream} requests in 
the memory controller read/write queues increases, \omcr{which in turn increases the} queuing delay for the
random read requests.\footnote{\atb{Note that the memory 
requests are \emph{only} subject to delays incurred at the memory controller 
because our simulation injects read and write requests directly into the memory
controller read and write queues.}}} 

\atb{We conclude that Ramulator~2.0, when configured and used correctly, yields reasonable \omcr{and realistic}
\nbcr{5}{bandwidth--latency} curves that resemble the ones obtained from real systems by~\cite{mess} \omcr{(see Section 3, Fig. 3 in~\cite{mess})}. The results \omcr{we present in this section}
\omtwo{thus} show that the Ramulator~2.0 results presented in~\cite{mess} are \omcr{incorrect} and are \emph{not}
representative of Ramulator~2.0's modeling accuracy \omcr{and capability}.}
\section{Evaluation of DAMOV in the \omtwo{Mess} Paper}
\label{sec:other-issues}
\subsection{\omtwo{Mess} Paper's \cite{mess} DAMOV (ZSim+Ramulator) Results}

\begin{figure}[!h]
    \centering
    \includegraphics[width=1\linewidth]{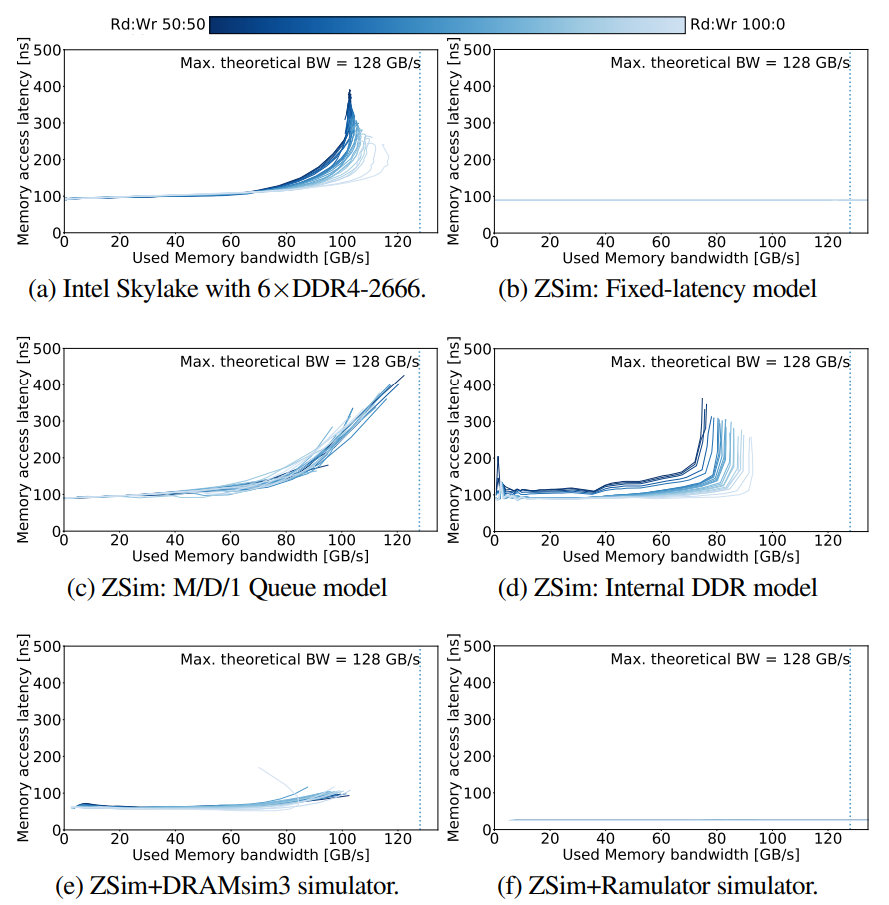}
    \caption{\atb{\nbcr{5}{Bandwidth--latency}} curves for ZSim memory models from the \omtwo{Mess} paper~\cite{mess} \omcr{(\omtwo{same as} Fig. 5 in~\cite{mess}}\agy{)}}
    \label{fig:mess_zsim}
\end{figure}

\hluo{Fig.~\ref{fig:mess_zsim} shows the \nbcr{5}{bandwidth--latency} curves of \atb{various} ZSim~\cite{zsim} memory models (sub-figures b, c, d, e, and f), compared to real system measurements (sub-figure a), from the \omtwo{Mess} paper~\cite{mess} (Fig. 5 in~\cite{mess}). The \omtwo{Mess} paper claims DAMOV (i.e., ZSim+Ramulator)~\cite{damov} ``provides a fixed 25 ns latency in the whole bandwidth area and for all memory traffic configurations'' (\omtwo{Section 4} in~\cite{mess}).}

\hluo{To understand why the \omtwo{Mess} paper shows \omcr{such} an unrealistic \omcr{(i.e., low and \omtwo{seemingly} constant)} \atb{memory} latency for DAMOV, we carefully checked the code and scripts from the artifact repository of the \omtwo{Mess} paper~\cite{messzenodo, messgit,messsimgit}. \om{We identify two major issues: \omcr{1)~irreproducible simulation, and 2)~erroneous memory latency statistics usage, which we describe in Sections~\ref{sec:irreproducible-sim} and~\ref{sec:erroneous-memory}}}}.

\subsection{\revision{Reproducibility Issues in \cite{mess}}}\rcomment{D1}
\label{sec:irreproducible-sim}
We carefully checked the artifact repositories of the \omtwo{Mess} paper\cite{messgit,messzenodo,messsimgit} based on the description of the artifact in Appendix C of~\cite{mess}. Unfortunately, we could \emph{\omcr{not}} find \omcr{the following three items used by the \omtwo{Mess} paper to produce the results in Fig. 5 (f) in the \omtwo{Mess} paper:} 1) the source code of the ZSim simulator, 2) the source code of Ramulator, and 3) the configuration of Ramulator used. We also checked the path to the simulator executable in the experiment scripts from the \omtwo{Mess} paper artifact~\cite{messzenodo,messgit,messsimgit}. Unfortunately, the \emph{hardcoded} path\footnote{\textttbreak{/home/bsc18/bsc18278/zsimdramsim3/acmSimulation/zsim.git\_mt/build/debug/zsim} from line 33 of \textttbreak{Mess-benchmark/CPU/Simulators/Execution-driven/ZSim/ramulator/submit.batch}} \omtwo{in~\cite{messgit,messzenodo}} points to a directory that does \emph{not} exist in the artifacts~\cite{messgit,messzenodo,messsimgit}.

We tried to independently reproduce the \omtwo{Mess} paper's results \omcr{(i.e., Fig. 5-f in~\cite{mess})} by combining the ZSim configuration available in the artifact repository of the \omtwo{Mess} paper with the Ramulator configuration from DAMOV~\cite{damov}. We modify the Ramulator configuration to model the $6\times$DDR4-2666 configuration from the \omtwo{Mess} paper in a best-guess manner. We perform the simulation using 1) the ZSim and Ramulator code from DAMOV~\cite{damov}, and 2) the \omtwo{Mess} paper's \texttt{stream\_mpi} workload. However, doing so triggers a rigid assertion in DAMOV because it does not support non-power-of-2 numbers of DRAM channels. 
\omcr{Therefore, we} cannot independently reproduce the DAMOV results in Fig. 5-f of the \omtwo{Mess} paper \omtwo{using
the configuration that the Mess paper describes that it used
to produce Fig.~5-f in~\cite{mess}.}

\subsection{\revision{Technical Simulator Usage Errors in \cite{mess}}}\rcomment{D1}
\label{sec:erroneous-memory}
\hluo{We find that the \omtwo{Mess} paper uses two statistics from ZSim to calculate the average memory latency for DAMOV, \texttt{latGETnl} (i.e., the accumulative latency) and \texttt{mGETs} (i.e., the number of GET requests) from the L1-D cache.\footnote{From lines 162-163 in \textttbreak{Mess-benchmark/CPU/Simulators/Execution-driven/ZSim/ramulator/processing/calculator.py}} These are \emph{not} the correct statistics to compute the memory latency for DAMOV because the bound-weave simulation methodology of ZSim~\cite{zsim} 1) only updates \texttt{latGETnl} with constant latencies (i.e., contention-free latency) throughout the memory hierarchy during the bound phase, and 2) only updates the \emph{core clock cycle} (instead of \texttt{latGETnl}) with the actual simulated DRAM latency from Ramulator during the weave phase. In other words, the 
\texttt{latGETnl} statistics from ZSim does \emph{not} \omtwo{include} any actual DRAM latency simulated by Ramulator, but only \omtwo{includes} the \emph{constant} contention-free latenc\omtwo{y}.}

\hluo{We find that the average memory latency for DAMOV as the \omtwo{Mess} paper measures (i.e., \texttt{latGETnl} / \texttt{mGETs}) is about 48 cycles. This matches closely with the accumulative contention-free latency from L1-D to the memory controller of 52 cycles (i.e., about 25ns for 2.1GHz CPU frequency) based on the ZSim configuration\footnote{From \textttbreak{Mess-benchmark/CPU/Simulators/Execution-driven/ZSim/ramulator/sb.cfg}} of the \omtwo{Mess} paper~\cite{mess}.}

\hluo{The DRAM performance metrics should come from Ramulator statistics output itself. Unfortunately, \emph{all} the Ramulator statistics output\footnote{E.g., \textttbreak{Mess-benchmark/CPU/Simulators/Execution-driven/ZSim/ramulator/measuring/bw-lat/measurment_100_0/bwLatCurves.ramulator.stats}} in the \omtwo{Mess} artifact are \emph{empty}.}
\omcr{We therefore believe that the results in the \omtwo{Mess} paper for DAMOV are 
reflecting incorrect statistics due to simulator usage error. }
\shepherd{Our findings demonstrate the need to 1) carefully configure, understand, and use simulators, and 2) validate findings before making strong claims about the accuracy of the simulators.}\scomment{16}{}

\subsection{\nb{Our Evaluation Methodology}}

\nb{To show that DAMOV (ZSim+Ramulator), when used correctly, yields reasonable \nbcr{5}{bandwidth--latency} curves and non-constant DRAM latency, we evaluate a modern DDR4 memory system in DAMOV using the Mess benchmark's streaming access workload and the ZSim configuration files in the Mess paper's artifact~\cite{messgit}.
Table~\ref{tab:damov_config} describes our system simulation configuration. We make three modifications to the configuration. First, to bypass the assertion failure that is described in Section~\ref{sec:irreproducible-sim}, we instantiate DRAM with 8 channels. This configuration matches the memory system of another real DDR4-based system that the Mess paper evaluates, IBM Power9~\cite{mess}. Second, we add the timing parameters of the DDR4-2666T standard~\cite{jedec2012ddr4} to remain consistent with what is evaluated in the Mess paper~\cite{mess}. Third, we revise the Mess paper's configuration file to avoid runtime errors (caused by configurations that DAMOV's open source version does \emph{not} support~\cite{damov}).}

\nb{To construct each \nbcr{5}{bandwidth--latency} curve, we run the simulation for 100 \nbcr{2}{million} instructions. We sweep NOP frequency from 1 to 750 and the read ratio from 50 to 100 \atb{to vary system load}. We report the memory access latency for read requests using Ramulator statistics. This enables us to measure the memory access latency at the memory controller level, excluding the delays caused by the core model and the cache hierarchy.}\footnote{\nb{To measure the end-to-end latency of a memory access, the ZSim core model needs to be modified to collect this statistic.}}

\vspace{1mm}

\begin{table}[ht]
\centering
\caption{Simulated System Configuration in DAMOV}
\begin{tabularx}{\linewidth}{lp{5cm}}
\toprule
\textbf{Processor:} & 24 cores @ 2.4~GHz\\
\midrule
\makecell[l]{\textbf{Cache Hierarchy:}} & L1i and L1d: 8-way 64~KB,
L2: 16-way 1~MB, LLC: 11-way 33~MBs.\\
\midrule
\textbf{DRAM:} & DDR4-2666T~\cite{jedec2012ddr4}; 8~channels \revision{(64-bit each)}, 1~rank, 4~bank~groups, 4~banks\\
\midrule
\makecell[l]{\textbf{Timing} \textbf{Parameters:}} & $t_{RCD}$: \SI{12.75}{\nano\second}, $t_{RAS}$: \SI{32.25}{\nano\second}, $t_{RP}$: \SI{12.75}{\nano\second}, $t_{RTP}$: \SI{7.5}{\nano\second},  $t_{CCD\_S}$: 4 nCK, $t_{RFC}$: \SI{261}{\nano\second}, $t_{REFI}$: \SI{7.8}{\micro\second}\\
\midrule
\makecell[l]{\textbf{Memory} \textbf{System:}} & FR-FCFS request scheduling policy~\cite{scott2000memory, zuravleff1997controller}, All-bank refresh, Read queue size:~32, Write queue size:~32 \\
\midrule
\end{tabularx}
\label{tab:damov_config}
\end{table}
\rcomment{C2}

\subsection{\nb{Key Results}}
\label{sec:damov_results}

\begin{figure}[!h]
    \centering
    \includegraphics[width=1\linewidth]{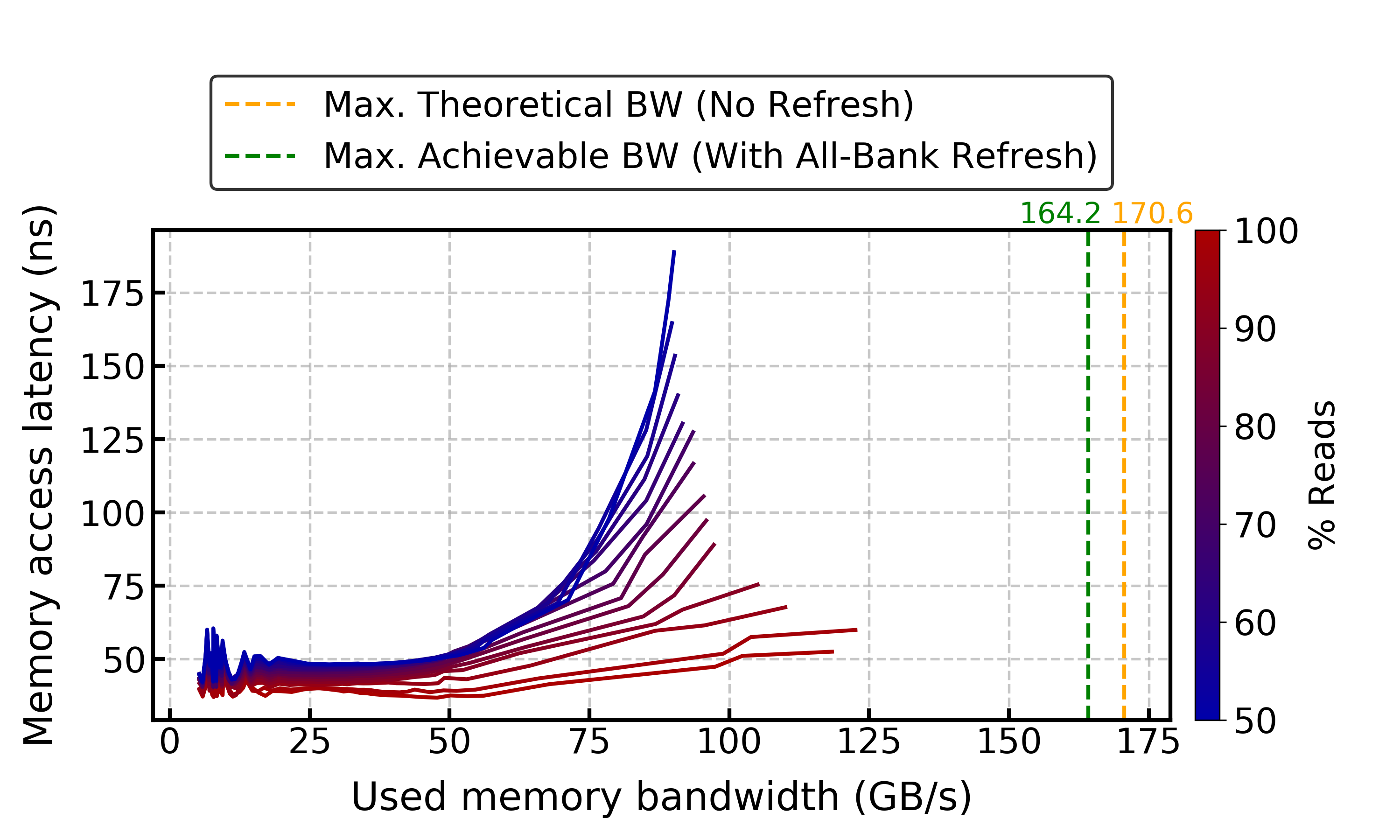}
    \caption{\atb{\nbcr{5}{Bandwidth--latency}} curves for DDR4-2666T in {DAMOV} using the configuration in Table~\ref{tab:damov_config}.}
    \label{fig:mess_damov}
\end{figure}

\nb{Fig.~\ref{fig:mess_damov} shows the \nbcr{5}{bandwidth--latency} curves generated using DAMOV with the configuration given by Table~\ref{tab:damov_config}. The maximum theoretical bandwidth for 8 channels DDR4-2666T (indicated by the orange line in the figure) is {170.6} GB/s. The maximum bandwidth \textit{cannot} be sustained during standard operation due to \emph{all-bank periodic refresh operations}~\cite{jedec2012ddr4} that \omcr{conflict with and delay} 
data transfers. 
\omcr{Taking such refresh operations and their impact on data transfers into account,} we compute the \emph{maximum achievable bandwidth}
from the maximum theoretical bandwidth
as {164.2} GB/s (indicated by the green line in the figure) from the $BW_{achievable}$ equation in~\secref{sec:results}.}

\nb{From Fig.~\ref{fig:mess_damov}, we observe that DAMOV's memory access latency is not constant, in contrast to what is reported in the Mess paper. As system load increases, the memory access latency increases. We conclude that when correct DRAM statistics are used, DAMOV's simulated memory access latency is \textit{not} constant.}
\section{\revision{Discussion}}\rcomment{C1,\\D2}
\label{sec:discussion}

\revision{For Ramulator~2.0 and DAMOV users to avoid the configuration and usage errors described 
in detail in this work, we provide four best practices for Ramulator~2.0 and DAMOV, and we integrate 
memory bandwidth statistics as a sanity check into {our open sourced version of}
Ramulator~2.0.
%'s open source repository.
}

\subsection{\revision{Ramulator~2.0 and DAMOV Best Practices}}

\takeaway{We recommend that Ramulator~2.0 users familiarize themselves 
with JEDEC standard documents (e.g., the JEDEC DDR5 \nbcr{3}{DRAM} standard~\cite{jedecddr5c}) 
instead of relying on 
\omthree{information other than the official standard or the simulator documentation.}
}
\revision{
\omthree{Importantly, users should be aware that colloquial technical terms may have different meanings compared to precise definitions in official standards and simulator implementations.}
\takeaway{Users should understand Ramulator~2.0's memory organization and configurations, which are 
\nb{documented}
in the code \nbcr{4}{and the repository}.}
The first Ramulator~2.0 configuration error (see~\ref{sec:mess-paper-errors}) in the Mess paper~\cite{mess} could have been avoided with the first two best practices.}

\takeaway{Users should understand simulation methodology
and simulator implementations sufficiently such that they use
the correct statistics.}
\revision{
The 
third best practice
\omthree{could} have {prevented}
{DAMOV simulator usage errors (see~\ref{sec:erroneous-memory})}
{in~\cite{mess}.}}

\takeaway{To ensure correct simulator usage, we recommend 
communicating with the simulator developers {or user community} when unexpected results are observed (in line with 
the open source spirit {of free software development~\cite{lakhani2003open}}).}

\nbcr{3}{All configuration and usage issues in~\cite{mess} that are described in our paper could have been avoided with the fourth best practice.}

\subsection{\revision{Memory Bandwidth Statistics as a Sanity Check}}

\revision{Ramulator~2.0 outputs standard statistics that
are useful for performance evaluation and debugging. 
To allow Ramulator~2.0 users to \omthree{more} easily determine 
if their simulation is acting unexpectedly 
with respect to the \omthree{\textit{maximum}} and \omthree{\textit{used}} {memory} bandwidth,
we add two new statistics to {our open sourced version of} Ramulator~2.0.
First, the \omthree{\textit{maximum bandwidth}} statistic
quantifies the maximum theoretical {memory bandwidth} that can
be attained by the simulated DRAM configuration ($BW_{theoretical}$ in~\secref{sec:results}).
Second, the \omthree{\textit{used {memory} bandwidth}} statistic quantifies the DRAM throughput
attained by the simulated workload (e.g., a point 
along a curve in Fig.~\ref{fig:main_res}).}
\section{Conclusion}

\omcr{We demonstrated that the claims made by 
the \omtwo{Mess} paper about Ramulator~2.0~\cite{ramulator2} and DAMOV~\cite{damov} are incorrect
and are due to \omtwo{simulator} configuration \omtwo{and simulator usage} errors made in the \omtwo{Mess} paper~\cite{mess}.}
\shepherd{We strongly encourage the computer architecture community to consider our corrections to the Ramulator 2.0 and DAMOV results of the Mess paper~\cite{mess} to prevent the propagation of inaccurate and misleading results and to maintain the reliability of the scientific record.} \scomment{17}{}
More broadly, 
simulators should be configured, \omtwo{understood,} and used very carefully \omtwo{and rigorously,} especially before making claims about them being wrong. 
\omtwo{The open source spirit also suggests that the authors of freely available simulators are properly
contacted to ensure these simulators are used with correct configurations and as intended, especially before publishing strong and incorrect claims about such simulators.}
\shepherd{Our investigation also aims to {stimulate discussion} on artifact evaluation practices and on mechanisms for correcting results and artifacts after publication.} \scomment{18}{}\rcomment{CQ2}
\nbcr{2}{To facilitate transparency and
reproducibility, we open source all our code and results at \githubLink{}.}

\section*{Acknowledgments} 

\nbcr{1}{We thank the anonymous reviewers of {ISPASS 2026} for feedback. 
\atbcr{1}{The anonymous reviewers heavily scrutinized this paper and provided
numerous suggestions, \nbcr{6}{in some cases} sentence by sentence, to ensure that each statement 
in this paper is correct and firmly backed up by data. If the strict scrutiny applied to this paper 
had been consistent across the scientific review process, 
we believe the Mess paper would \emph{not} have been published in its current state.}
We believe our results and analyses in this paper open up broader \nbcr{2}{questions regarding} the review and artifact \nbcr{2}{evaluation} processes.
\nbcr{6}{In particular, it is critical for tool papers that claim incorrect results in freely available open-source software to be held to the standard of being asked to have checked with the authors and maintainers of the software first, to establish the accuracy of their results.}
We thank practitioners from 
\nbcr{2}{many major companies in industry and researchers \nbcr{3}{at} many institutions in academia}
who \nbcr{2}{have been using} Ramulator and \nbcr{2}{providing} valuable feedback 
since the release of the first \nbcr{2}{version in 2015, as well as the second version in 2023}. Their experiences and comments helped \nbcr{3}{us} validate the \nbcr{2}{practicality, usability, and accuracy}
of our tools.
\nbcr{6}{Some of these folks have also read the Mess paper and expressed their surprise that it was selected as a "best paper runner up" and even accepted, even though its authors had no communication with the developers and maintainers of Ramulator~2.0.}
{We thank the} SAFARI Research Group members for
{constructive} feedback and the stimulating \nbcr{3}{\& open} intellectual \nbcr{3}{and scientific} {environment \nbcr{3}{they provide}.}
We acknowledge the generous gift funding provided by our industrial partners
({especially} Google, Huawei, Intel, Microsoft), which has been instrumental in
enabling the \nbcr{3}{extensive} research we have been conducting on memory systems \nbcr{5}{for more than two decades}~\cite{mutlu2020modern,mutlu2013memory,mutlu2025memory,mutlu2019rowhammer,mutlu2023fundamentally,mutlu2019processing,cai2017flashtbd,mutlu2014research,singh2021fpga,mutlu2017rowhammer,mutlu2020intelligentdate,oliveira2022accelerating,mutlu2024memory}. This work was in part
supported by \nbcr{2}{a} Google Security and Privacy Research Award and the Microsoft
Swiss Joint Research Center.}

% \balance

%%%% this is from the IEEE template
\bibliographystyle{unsrt}
\bibliography{refs}

@ARTICLE{ramulator1,
  title = {{Ramulator: A Fast and Extensible DRAM Simulator}},
  author = {Kim, Yoongu and Yang, Weikun and Mutlu, Onur},
  journal = {{IEEE CAL}},
  year = {2015},
}

@ARTICLE{ramulator2,
  author={Luo, Haocong and Tugrul, Yahya Can and Bostanci, F. Nisa and Olgun, Ataberk and Yaglikci, A. Giray and Mutlu, Onur},
  journal={IEEE CAL}, 
  title={{Ramulator 2.0: A Modern, Modular, and Extensible DRAM Simulator}}, 
  year={2024}}

@INPROCEEDINGS{mess,
  author={Esmaili-Dokht, Pouya and Sgherzi, Francesco and Girelli, Valéria Soldera and Boixaderas, Isaac and Carmin, Mariana and Monemi, Alireza and Armejach, Adrià and Mercadal, Estanislao and Llort, Germán and Radojković, Petar and Moreto, Miquel and Giménez, Judit and Martorell, Xavier and Ayguadé, Eduard and Labarta, Jesus and Confalonieri, Emanuele and Dubey, Rishabh and Adlard, Jason},
  booktitle={MICRO}, 
  title={{A Mess of Memory System Benchmarking, Simulation and Application Profiling}}, 
  year={2024}}

@misc{messgit,
  title={{Mess benchmark}},
  author={{Memory systems for HPC and AI @BSC}},
  year={2024},
  howpublished={\url{https://github.com/bsc-mem/Mess-benchmark}}
}

@misc{messsimgit,
  title={{Mess simulator}},
  author={{Memory systems for HPC and AI @BSC}},
  year={2025},
  howpublished={\url{https://github.com/bsc-mem/Mess-simulator}}
}

@misc{messzenodo,
  title={{A Mess of Memory System Benchmarking, Simulation and Application Profiling}},
  author={{Esmaili-Dokht, Pouya}},
  year={2024},
  howpublished={\url{https://zenodo.org/records/13748674}}
}

@inproceedings{verdejo2017microbenchmarks,
  title={{Microbenchmarks for Detailed Validation and Tuning of Hardware Simulators}},
  author={Verdejo, Rommel S{\'a}nchez and Radojkovic, Petar},
  booktitle={HPCS},
  year={2017},
}

@article{oliveira2021damov,
  title={{DAMOV: A New Methodology and Benchmark Suite for Evaluating Data Movement Bottlenecks}},
  author={Oliveira, Geraldo F and G{\'o}mez-Luna, Juan and Orosa, Lois and Ghose, Saugata and Vijaykumar, Nandita and Fernandez, Ivan and Sadrosadati, Mohammad and Mutlu, Onur},
  journal={IEEE Access},
  year={2021},
}

@inproceedings{verdejo2018main,
  title={{Main Memory Latency Simulation: The Missing Link}},
  author={Verdejo, Rommel S{\'a}nchez and Asifuzzaman, Kazi and Radulovic, Milan and Radojkovi{\'c}, Petar and Ayguad{\'e}, Eduard and Jacob, Bruce},
  booktitle={MEMSYS},
  year={2018}
}

@inproceedings{scott2000memory,
  title={Memory Access Scheduling},
  author={Rixner, Scott and Dally, William J. and Kapasi, Ujval J. and Mattson, Peter and Owens, John D.},
  booktitle={ISCA},
  year={2000}
}

@misc{zuravleff1997controller,
  title={{Controller for a synchronous DRAM that maximizes throughput by allowing memory requests and commands to be issued out of order}},
  author={Zuravleff, William K and Robinson, Timothy},
  year={1997},
  howpublished={US Patent 5,630,096}
}

@manual{jedecddr5c,
  title="{JESD79-5C: DDR5 SDRAM Standard}",
  author="{{JEDEC}}",
  year={2024}
}

@manual{jedecddr5udimm,
  title={{JESD308B: DDR5 Unbuffered Dual Inline Memory Module (UDIMM) Common Standard}},
  author={{JEDEC}},
  year={2025}
}

@manual{jedecddr5rdimm,
  title={{JESD305A: DDR5 Registered Dual Inline Memory Module (RDIMM) Common Standard}},
  author={{JEDEC}},
  year={2025}
}

@misc{ramulator1github,
	title        = {{Ramulator}},
	author       = {SAFARI Research Group},
	howpublished = {\url{https://github.com/CMU-SAFARI/ramulator}}
}

@misc{ramulator2github,
	title        = {{Ramulator 2.0}},
	author       = {SAFARI Research Group},
	howpublished = {\url{https://github.com/CMU-SAFARI/ramulator2}}
}

@misc{damov,
	title        = {{DAMOV: A New Methodology and Benchmark Suite for Evaluating Data Movement Bottlenecks}},
	author       = {SAFARI Research Group},
	howpublished = {\url{https://github.com/CMU-SAFARI/DAMOV}}
}

@inproceedings{zsim,
author = {Sanchez, Daniel and Kozyrakis, Christos},
title = {{ZSim: Fast and Accurate Microarchitectural Simulation of Thousand-core Systems}},
year = {2013},
booktitle = {ISCA}
}

@misc{jedec2012ddr4,
  title="{JESD79-4 DDR4 SDRAM standard}",
  author={JEDEC},
  year={2012}
}

@misc{lee2010dram,
  title={{DRAM-Aware Last-Level Cache Writeback: Reducing Write-Caused Interference in Memory Systems}},
  author={Lee, Chang Joo and Narasiman, Veynu and Ebrahimi, Eiman and Mutlu, Onur and Patt, Yale N},
  year={2010},
  howpublished={Technical Report HPS-2010-002}
}

@article{lakhani2003open,
  title={{How Open Source Software Works: ``Free''' User-to-user Assistance}},
  author={Lakhani, Karim R and Von Hippel, Eric},
  journal={Research policy},
  year={2003}
}

@misc{self.github,
	title        = {{Ramulator 2.0 -- Mess benchmark}},
	author       = {SAFARI Research Group},
	howpublished = {\url{https://github.com/CMU-SAFARI/ramulator2/tree/mess}}
}

@misc{cutmess.github,
	title        = {{Cleaning up the Mess Source Code}},
	author       = {SAFARI Research Group},
	howpublished = {\url{https://github.com/CMU-SAFARI/Cleaning-up-the-Mess}}
}

@incollection{mutlu2020modern,
	title        = {{A Modern Primer on Processing in Memory}},
	author       = {Mutlu, Onur and Ghose, Saugata and G{\'o}mez-Luna, Juan and Ausavarungnirun, Rachata and Sadrosadati, Mohammad and Oliveira, Geraldo F},
	year         = 2021,
	booktitle    = {Emerging Computing: From Devices to Systems --- Looking Beyond Moore and Von Neumann},
	publisher    = {Springer},
	url          = {https://arxiv.org/abs/2012.03112}
}

@inproceedings{mutlu2013memory,
	title        = {{Memory Scaling: A Systems Architecture Perspective}},
	author       = {Mutlu, Onur},
	year         = 2013,
	booktitle    = {IMW}
}

@inproceedings{mutlu2025memory,
  title={{Memory-Centric Computing: Solving Computing’s Memory Problem}},
  author={Mutlu, Onur and Olgun, Ataberk and Yuksel, Ismail Emir},
  booktitle={IMW},
  year={2025}
}

@article{mutlu2019rowhammer,
	title        = {{RowHammer: A Retrospective}},
	author       = {Mutlu, Onur and Kim, Jeremie S},
	year         = 2019,
	journal      = {TCAD},
	booktitle    = {COSADE}
}

@inproceedings{mutlu2023fundamentally,
	title        = {{Fundamentally Understanding and Solving RowHammer}},
	author       = {Mutlu, Onur and Olgun, Ataberk and Yaglikci, A. Giray},
	year         = 2023,
	booktitle    = {ASP-DAC}
}

@inproceedings{mutlu2019processing,
	title        = {{Processing Data Where It Makes Sense: Enabling In-Memory Computation}},
	author       = {Mutlu, Onur and Ghose, Saugata and G{\'{o}}mez-Luna, Juan and Ausavarungnirun, Rachata},
	year         = 2019,
	journal      = {MICPRO},
	booktitle    = {Microprocessors and Microsystems}
}

@inproceedings{mutlu2024memory,
  title={{Memory-Centric Computing: Recent Advances in Processing-in-DRAM}},
  author={Mutlu, Onur and Olgun, Ataberk and Oliveira, Geraldo F and Yuksel, Ismail E},
  booktitle={IEDM},
  year={2024}
}

@article{oliveira2022accelerating,
	title        = {{Accelerating Neural Network Inference with Processing-in-DRAM: From the Edge to the Cloud}},
	author       = {Oliveira, Geraldo F and G{\'o}mez-Luna, Juan and Ghose, Saugata and Boroumand, Amirali and Mutlu, Onur},
	year         = 2022,
	journal      = {IEEE Micro}
}

@inproceedings{mutlu2020intelligentdate,
	title        = {{Intelligent Architectures for Intelligent Computing Systems}},
	author       = {Mutlu, Onur},
	year         = 2021,
	booktitle    = {DATE}
}

@inproceedings{mutlu2017rowhammer,
	title        = {{The RowHammer Problem and Other Issues We May Face as Memory Becomes Denser}},
	author       = {Mutlu, Onur},
	year         = 2017,
	booktitle    = {DATE}
}

@article{singh2021fpga,
	title        = {{FPGA-Based Near-Memory Acceleration of Modern Data-Intensive Applications}},
	author       = {Singh, Gagandeep and Alser, Mohammed and Cali, Damla Senol and Diamantopoulos, Dionysios and G{\'o}mez-Luna, Juan and Corporaal, Henk and Mutlu, Onur},
	year         = 2021,
	journal      = {IEEE Micro}
}

@article{mutlu2014research,
	title        = {{Research Problems and Opportunities in Memory Systems}},
	author       = {Mutlu, Onur and Subramanian, Lavanya},
	year         = 2014,
	journal      = {SUPERFRI}
}

@article{cai2017flashtbd,
	title        = {{Error Characterization, Mitigation, and Recovery in Flash Memory Based Solid-State Drives}},
	author       = {Y. Cai and S. Ghose and E. F. Haratsch and Y. Luo and O. Mutlu},
	year         = 2017,
	journal      = {Proc. IEEE}
}

% \clearpage
\appendix
\setstretch{0.99}

\section{Artifact Appendix}

%%%%%%%%%%%%%%%%%%%%%%%%%%%%%%%%%%%%%%%%%%%%%%%%%%%%%%%%%%%%%%%%%%%%%
\subsection{Abstract}
This artifact provides the source code and scripts to reproduce the key experiments and their results presented in our ISPASS 2026 paper. This artifact enables reproducing the following key results and figures:

\begin{enumerate}
\item \nbcr{5}{Bandwidth--latency} curves for DDR5-4800AN in Ramulator 2.0 (Figures 1.c and 3)
\item \nbcr{5}{Bandwidth--latency} curves for DDR4-2666\atbt{T} in DAMOV (Figure 5)
\end{enumerate}

The artifact is built on two simulation environments: Ramulator 2.0 and DAMOV. These two simulation environments and the corresponding experiments are independent, and we intend them to run separately. 

\subsection{Artifact Check-List (Meta-Information)}

{\small
\begin{itemize}
    \item {\bf Program:} C++ programs, Python scripts, shell scripts.
    \item {\bf Compilation:} clang, cmake, GNU make, scons.
    \item {\bf Run-time environment: } Linux (tested on Ubuntu 20.04 and 22.04), Python 3, (optional) Docker.
    \item {\bf Execution: } Python scripts.
    \item {\bf Metrics: } Memory access latency (in ns), Used memory bandwidth (GB/s).
    \item {\bf Output: } Figures in PDF and PNG format and related results as script outputs.
    \item {\bf How much disk space required (approximately)?: } 5 GB.
    \item {\bf How much time is needed to prepare workflow (approximately)?: } 10-30 minutes.
    \item {\bf How much time is needed to complete experiments (approximately)?:} 1-2 hours, given enough compute resources \atbt{(e.g., approximately 200 CPU cores)}.
    \item {\bf Publicly available?: } Yes.
    \item {\bf Archived (provide DOI)?: } Yes, DOI: \url{https://doi.org/10.5281/zenodo.18911020}
\end{itemize}
}

%%%%%%%%%%%%%%%%%%%%%%%%%%%%%%%%%%%%%%%%%%%%%%%%%%%%%%%%%%%%%%%%%%%%%
\subsection{Description}

\subsubsection{How to Access} The source code and scripts can be downloaded from Zenodo (\url{https://doi.org/10.5281/zenodo.18911020}).

\subsubsection{Hardware Dependencies}
The artifact is designed for Linux-based systems and can be run either on \textbf{a Slurm-based cluster infrastructure} or on \textbf{a personal computer}. 

The full evaluation of our DAMOV results requires many multithreaded simulation runs, therefore, we strongly recommend using a Slurm-based environment \atbt{with many CPU cores} whenever possible.
To enable easy reproduction of our DAMOV results, we provide SSH access to our internal Slurm-based infrastructure, including all required hardware and software, during the artifact evaluation. Please contact us through HotCRP and/or the AE committee for details.

Our tested system specifications:
\begin{itemize}
    \item Docker environments with Linux 22.04 images assuming x86-64 systems
    \item (For DAMOV experiments) Supporting at least 48 GB memory and 24 threads per experiment
    \item (For Ramulator 2.0 experiments) Supporting at least 6 GB memory per single-threaded experiment
\end{itemize}

\subsubsection{Software Dependencies}
\begin{itemize}
    \item Linux Operating System (tested on Ubuntu 20.04 and 22.04).
    \item (Optional) Docker
    \item (Recommended) Slurm
    \item Python3 (version 3.8)
    \item Packages: pandas numpy matplotlib seaborn pyyaml scipy scons
    \item CMake, GNU Make 

\end{itemize}

All experiments are run with Python scripts:

\fancycommand{apt-get install python3-dev python3-pip}

Plotting packages can be installed with:

\fancycommand{pip3 install  matplotlib pandas numpy}

All requirements are installed with build scripts automatically:

\begin{itemize}
\item \textbf{DAMOV:} build-essential, scons, automake, autoconf, m4, perl, flex, bison, byacc, libhdf5-dev, libelf-dev, python3-dev, python3-pip, openmpi-bin, libopenmpi-dev.
\item \textbf{Ramulator 2.0:} argparse, spdlog, yaml-cpp.
\end{itemize}

\subsection{Installation and Experiment Workflow}

\subsubsection{Ramulator 2.0 Experiments}

Ramulator 2.0 experiments are intended to run on a personal computer using a Docker image that handles all dependencies and sets up the simulation environment.

\noindent\textbf{To build the simulator, run all experiments and plot inside the Docker container:}

\fancycommand{cd ramulator2 \\
./build\_in\_docker.sh}

This will build ramulator2 from source inside an Ubuntu 22.04 container, run all simulation experiments (\texttt{run.py}), generate the plots (\texttt{plot.py}), and copy the figures (\texttt{latency\_bandwidth\_16ch.pdf} and \texttt{latency\_bandwidth\_16ch.png}) and figures back to the \texttt{mess/} directory.

Alternatively, you can build and run Ramulator 2.0 natively by following the instructions in the \texttt{ramulator2/README.md} file.

\noindent\textbf{Customizing Evaluation Setup.}
You can configure the maximum number of threads to concurrently run the experiments \atbt{in} \texttt{ramulator2/mess/run.py}

\fancycommand{
... \\
\# Number of concurrent threads \\
MAX\_WORKERS = 16 \\
...
}

\subsubsection{DAMOV Experiments}

DAMOV experiments are intended to run on a Slurm-based infrastructure using a native build or a Docker image. Our scripts satisfy all requirements and run experiments automatically.

{Note that during the dependency setup phase, the script first tries to set up libraries with \texttt{apt-get} command, which can yield error outputs when the user does not have the privilege to execute this command. If this happens, the script falls back to setting up libraries locally by downloading the packages and copying them to the appropriate directories (\texttt{/simulator/libconfig}).}

\noindent\textbf{To set up dependencies, compile simulators and workloads, and run experiments in native environment:}

\fancycommand{
cd DAMOV/ \\
python3 run\_artifact.py
}

\noindent\textbf{(Optional) To use a Docker container:}

\fancycommand{
cd DAMOV/ \\
python3 run\_artifact\_container.py
}

This will build DAMOV and workloads from source, and schedule 700+ Slurm jobs. 

\noindent\textbf{(Optional) For a fast verification with fewer experiments: }

\fancycommand{
\# native: \\
python3 run\_artifact.py --fast \\  
\# Docker container \\
python3 run\_artifact\_container.py --fast
}

This will build DAMOV and workloads from source, and schedule fewer Slurm jobs (e.g., 10-20).

\noindent\textbf{Customizing Evaluation Setup.}
Please see the \texttt{DAMOV/README.md} file for instructions to further customize the evaluation setup.

\noindent\textbf{Plotting the Results}
After all experiments are complete, you can generate Figure 5 from the paper.

\fancycommand{python3 
\\
simulator/scripts/plot\_latency\_bw\_curve.py}

This will generate the figures (\texttt{figure5.png} and \texttt{figure5.pdf}) in the \texttt{DAMOV/} directory.

%%%%%%%%%%%%%%%%%%%%%%%%%%%%%%%%%%%%%%%%%%%%%%%%%%%%%%%%%%%%%%%%%%%%%
\subsection{Evaluation and Expected Results}

\subsubsection{Ramulator 2.0 Results}
Figure 1.c and 3 are expected to be reproduced as \texttt{ramulator2/mess/latency\_bandwidth\_16ch.pdf}. We expect the run script to reproduce \atbt{the two} figures and output the maximum bandwidth reported in our paper.

\subsubsection{DAMOV Results}
Figure 5 is expected to be reproduced as \texttt{DAMOV/figure5.pdf}. We expect the run script to reproduce the figure and output the maximum bandwidth reported in our paper.
Please note that DAMOV experiments include multithreaded experiments, which result in small changes in the produced curves and the maximum bandwidth value \atbt{across experiment runs}. However, the conclusions of our paper still apply to the reproduced figures and results, as the trends remain the same.

%%%%%%%%%%%%%%%%%%%%%%%%%%%%%%%%%%%%%%%%%%%%%%%%%%%%%%%%%%%%%%%%%%%%%

\subsection{Troubleshooting}
To retry failing Slurm jobs (for any reason), the run script (e.g., \texttt{DAMOV/run\_artifact.py}) can be run again. By default, the script skips completed experiments and only resubmit/rerun an experiment if it failed previously. Note that the run script should be executed \textit{after} all remaining jobs are completed to avoid disrupting existing jobs.

%%%%%%%%%%%%%%%%%%%%%%%%%%%%%%%%%%%%%%%%%%%%%%%%%%%%%%%%%%%%%%%%%%%%%
\subsection{Methodology}

Submission, reviewing, and badging methodology:

\begin{itemize}
  \item \url{https://www.acm.org/publications/policies/artifact-review-and-badging-current}
  \item \url{https://cTuning.org/ae}
\end{itemize}

\section{\nbcr{2}{Evaluating the Mess Benchmark\\on \nbcr{4}{Standalone} Ramulator}}

\nbcr{3}{The Mess paper~\cite{mess} evaluates the memory access latency and the used memory bandwidth behavior of DAMOV (i.e., ZSim+Ramulator) running the Mess benchmark (Fig. 5 in~\cite{mess}) and \nbcr{4}{claims} that DAMOV "provides a fixed 25 ns latency in the whole bandwidth area and for all memory traffic configurations" (Section 4 in~\cite{mess}). After careful analysis of the artifact repositories of the Mess paper~\cite{mess, messgit, messzenodo}, we find that the Mess paper uses two L1-D statistics \nbcr{4}{(i.e., \texttt{latGETnl} and \texttt{mGETs})} from ZSim to calculate the average memory latency for DAMOV, and \textit{not} the correct DRAM statistics provided by Ramulator. We independently evaluate DAMOV's simulation accuracy in~\secref{sec:damov_results} and show that when correct simulation statistics are used, DAMOV's memory access latency is \nbcr{4}{\emph{not}} constant (\figref{fig:mess_damov} \nbcr{4}{in~\secref{sec:damov_results}}).}

%DAMOV's memory access latency and curves (as shown in~\figref{fig:mess_damov}) more closely resemble the real system performance than the constant latency curves reported in the Mess paper~\cite{mess}. 

\nbcr{3}{Our results \nbcr{4}{from this investigation} (\figref{fig:mess_damov} \nbcr{4}{in~\secref{sec:damov_results}}) depict}
%However, the results show 
a gap between the maximum used memory bandwidth (shown as the rightmost endpoint of the red curves) and the maximum achievable bandwidth (shown as the green line).
\nbcr{3}{In this section,} we investigate this bandwidth gap to better understand the underlying simulated behavior better. 

% \nbcrcomment{moved this up}
We hypothesize that the bandwidth gap \nbcr{4}{is} caused by a combination of workload characteristics, system configuration, and \nbcr{4}{the integration of the core and memory models}. DAMOV executes the Mess benchmark by compiling and running the benchmark resources provided in the Mess paper’s artifact~\cite{messgit}. This is inherently different from crafting the memory access pattern of the Mess benchmark with a dedicated frontend module \nbcr{3}{(similar to our  Ramulator 2.0 evaluation methodology described in~\secref{sec:evaluation_ramulator})}. The memory access pattern of the workload executed on the simulated core model depends on many factors that might specifically hurt the streaming access pattern (e.g., DRAM address mapping) or yield a memory access frequency that is not sufficiently stressing the memory (e.g., core model optimizations). Therefore, this setup may not yield a memory access pattern that \nbcr{4}{fully utilizes} all available memory bandwidth.

To determine the \nbcr{4}{cause of} this gap and whether it is \nbcr{4}{due to} the memory model of Ramulator~\cite{ramulator1,ramulator1github}, we evaluate a DDR4 memory system in Ramulator using the same access pattern as described in the Mess benchmark.
\nbcr{2}{Note that the Ramulator version used in DAMOV is \textit{not} Ramulator~2.0 and is an earlier \nbcr{3}{Ramulator} version open-sourced at~\cite{ramulator1github}.} 
Similar to our evaluation methodology for our Ramulator~2.0 results (\secref{sec:evaluation_ramulator}), we introduce a new Ramulator frontend module, called \textit{MessProcessor}. The MessProcessor frontend is designed to emulate the memory access patterns of the Mess benchmark by generating a mixture of random access and streaming read requests, similar to the Mess Request Generator for Ramulator~2.0. \nbcr{3}{Using the MessProcessor frontend, \nbcr{5}{standalone} Ramulator performs memory accesses with the patterns of the Mess benchmark generated directly within Ramulator. In contrast, DAMOV executes the Mess benchmark (from the artifact repository of the Mess paper~\cite{cutmess.github}), and the memory requests are generated by the ZSim cores and fed to Ramulator's memory system.}

We use the MessProcessor frontend to construct the \nbcr{5}{bandwidth--latency} curves \nbcr{3}{for a DDR4 system}. We use the Ramulator memory system configuration in Table~\ref{tab:ramulator_config}. We open source our Ramulator configuration, along with all source code and scripts to reproduce our results~\cite{cutmess.github}.

\begin{table}[h]
\centering
\caption{Simulated System Configuration for \nbcr{4}{Standalone} Ramulator}
\begin{tabularx}{\linewidth}{lp{5cm}}
\toprule
\textbf{Frontend:} & \omtwo{Mess}Processor for Ramulator~\cite{ramulator1}.\\
\midrule
\textbf{DRAM:} & DDR4-2666T~\cite{jedec2012ddr4}; 8~channels \revision{(64-bit each)}, 1~rank, 4~bank~groups, 4~banks\\
\midrule
\makecell[l]{\textbf{Timing} \textbf{Parameters:}} & $t_{RCD}$: \SI{12.75}{\nano\second}, $t_{RAS}$: \SI{32.25}{\nano\second}, $t_{RP}$: \SI{12.75}{\nano\second}, $t_{RTP}$: \SI{7.5}{\nano\second},  $t_{CCD\_S}$: 4 nCK, $t_{RFC}$: \SI{261}{\nano\second}, $t_{REFI}$: \SI{7.8}{\micro\second}\\
\midrule
\makecell[l]{\textbf{Memory} \textbf{System:}} & FR-FCFS request scheduling policy~\cite{scott2000memory, zuravleff1997controller}, All-bank refresh, Read queue size:~32, Write queue size:~32 \\
\midrule
\end{tabularx}
\label{tab:ramulator_config}
\end{table}

\begin{figure}[!h]
    \centering
    \includegraphics[width=1\linewidth]{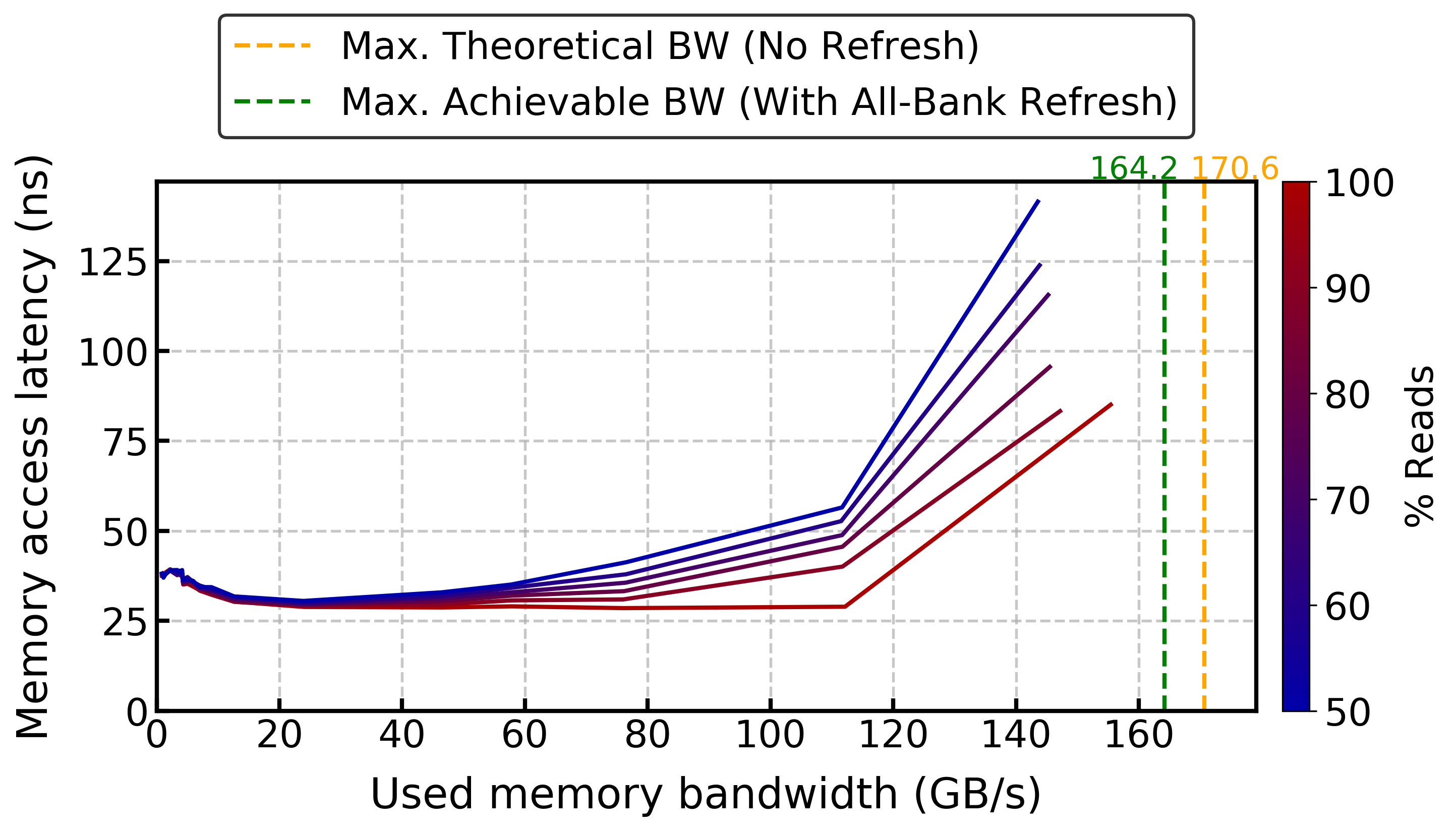}
    \caption{\nbcr{5}{Bandwidth--latency} curves for DDR4-2666T in \nbcr{4}{standalone} Ramulator~\cite{ramulator1, ramulator1github} using the \omtwo{MessProcessor} frontend \omcr{and configuration in Table~\ref{tab:ramulator_config}}}
    \label{fig:ramulator_old_mess}
\end{figure}

\figref{fig:ramulator_old_mess} shows the \nbcr{5}{bandwidth--latency} curves generated using \nbcr{5}{standalone} Ramulator with the MessProcessor frontend and the configuration given by Table~\ref{tab:ramulator_config}. The maximum theoretical bandwidth (yellow line) and the maximum achievable bandwidth (green line) are calculated as described in~\secref{sec:damov_results}.

We make two key observations from~\figref{fig:ramulator_old_mess}. First, the maximum used memory bandwidth (158.9 GB/s for \omcr{a read ratio of 1, i.e.,} 100\% reads, and NOP \omcr{frequency of 1, i.e., no NOP cycles}) is close to the maximum achievable bandwidth (164.2 GB/s). We attribute the remaining difference between the values to the random accesses in the Mess benchmark access pattern. Achieving maximum bandwidth requires exploiting row buffer locality and all levels of parallelism (e.g., bank-level and bank-group-level). \nbcr{4}{Random accesses in the Mess benchmark access pattern can} introduce row buffer conflicts and reduce the bank and bank-group-level parallelism. 

Second, the maximum used bandwidth reported by \nbcr{4}{the standalone} Ramulator \nbcr{3}{(with the MessProcessor frontend)} is higher than the maximum used bandwidth reported by DAMOV \nbcr{3}{(i.e., ZSim+Ramulator)}. This indicates that the bandwidth gap between DAMOV's maximum used bandwidth and the maximum achievable bandwidth is \textit{not} caused by the memory model of Ramulator. 

We conclude that \nbcr{4}{the standalone} Ramulator yields reasonable \nbcr{5}{bandwidth--latency} curves with a higher maximum used bandwidth value compared to DAMOV, and thus, the bandwidth gap between \nbcr{4}{the maximum used bandwidth} and the maximum achievable bandwidth is \textit{not} due to Ramulator's memory model.

% We hpyothesize that a combination of the workload, the system configuration and integration results in the having a larger gap between the maximum used bandwidth and the maximum achievable bandwidth. 
% First, DAMOV executes the Mess benchmark produced by compiling the benchmark resources in the Mess paper's artifact. Combined with the simulated system configurations regarding the DRAM address mapping and zsim and Ramulator integration, this might not directly translate into a memory access pattern that exploits row buffer locality and bank and bank group-level parallelism. 

\end{document}